\documentclass[twocolumn,superscriptaddress,amsmath,amssymb]{revtex4}
\usepackage{amsmath}
\usepackage{amssymb}
\usepackage{babel}
\usepackage{graphicx}% Include figure files
\usepackage{dcolumn}% Align table columns on decimal point
\usepackage{bm}% bold math
\usepackage{hyperref}% add hypertext capabilities
\usepackage{mathrsfs} % pour le P de la partie principale
\usepackage{gensymb} % symbole degré
\usepackage[absolute,overlay]{textpos}
\usepackage{upgreek} %grecques pour um et pour beta-decay
\usepackage{textcomp} % symbole pour mille

\date{June 2021}

\begin{document}

\title{Elasto-Raman scattering: Arsenic optical phonon as a probe of nematicity in BaFe$_2$As$_2$}

\author{Jean-Côme Philippe}
\email{jeancome.philippe@u-paris.fr}
\author{Jimmy Faria}
\affiliation{Université de Paris, Matériaux et Phénomènes Quantiques, UMR CNRS 7162, Bâtiment Condorcet, 75205 Paris Cedex 13, France}
\author{Anne Forget}
\author{Dorothée Colson}
\affiliation{Service de Physique de l'Etat Condensé, DSM/DRECAM/SPEC, CEA Saclay, Gif-sur-Yvette, 91191, France}
\author{Sarah Houver}
\author{Maximilien Cazayous}
\author{Alain Sacuto}
\affiliation{Université de Paris, Matériaux et Phénomènes Quantiques, UMR CNRS 7162, Bâtiment Condorcet, 75205 Paris Cedex 13, France}
\author{Yann Gallais}
\email{yann.gallais@u-paris.fr}
\affiliation{Université de Paris, Matériaux et Phénomènes Quantiques, UMR CNRS 7162, Bâtiment Condorcet, 75205 Paris Cedex 13, France}

\date{\today}

\begin{abstract}
We report a Raman scattering study of nematic degrees of freedom in the iron-based superconductor parent compound BaFe$_2$As$_2$ under tunable uniaxial strain. We demonstrate that the polarization resolved arsenic (As) phonon intensity can be used to monitor the nematic order parameter as a function of both temperature and strain. At low temperature in the nematic ordered phase we use it to track the continuous and reversible orientation of nematic domains under variable strain. At higher temperature, the evolution of the As phonon intensity under strain reflects an enhanced nematic susceptibility close to the nematic transition $T_S$. Its temperature dependence under strong strain follows qualitatively the expected behavior of an Ising order parameter under a symmetry breaking field. Our elasto-Raman study illustrates the interest of combining selective anisotropic strain with a symmetry resolved probe like Raman scattering. Elasto-Raman scattering can be applied to a wide variety of quantum materials where uniaxial strain tunes electronic orders.
\end{abstract}

\maketitle

\section{Introduction}
%discuss strain in Fe SC and quantum materials in general: new control parameter and way to probe nematic OP (in nematic SC like Fe SC)
% Need to design probes under continuous strain beyond transport: ARPES, optics, thermodynamics, NMR, STM, Xray.
Anisotropic strain is emerging as a valuable non-thermal tuning parameter of quantum materials \cite{chu_divergent_2012,hicks_strong_2014,steppke_strong_2017,kim_uniaxial_2018}. In the case of iron-based superconductors (Fe SC) where a nematic phase breaking the $C_4$ rotational symmetry of the underlying lattice is found in close proximity of superconductivity \cite{fernandes_what_2014}, uniaxial strain acts as a conjugate field of the nematic order parameter. It is therefore a useful tool to study nematic susceptibilities like elasto-resistivity \cite{chu_divergent_2012,hosoi_nematic_2016,kuo_ubiquitous_2016}, the change in resistivity anisotropy under uniaxial strain, but also more generally to study the interplay between nematic and superconducting orders \cite{malinowski_suppression_2020}. Moreover applying an uniaxial strain whose symmetry is transverse to the one of the nematic order parameter can provide an alternative path to tune a material towards a nematic quantum critical point with enhanced quantum critical fluctuations \cite{maharaj_transverse_2017,worasaran_nematic_2021}.
\par
Initially limited to transport measurements \cite{chu_divergent_2012,kuo_measurement_2013,hicks_piezoelectric-based_2014,hicks_strong_2014,steppke_strong_2017,bartlett_relationship_2021}, techniques under tunable uniaxial strain have been recently implemented and leveraged to study several superconducting materials, with in the case of Fe SC a particular focus on accessing nematic degrees of freedom \cite{mirri_origin_2015, mirri_electrodynamic_2016,kissikov_nuclear_2017,kissikov_uniaxial_2018,andrade_visualizing_2018,pfau_momentum_2019,sanchez_spontaneous_2020,hristov_elastoresistive_2019,ikeda_elastocaloric_2020,caglieris_strain_2021}. However few of these techniques are symmetry resolved and as such they do not directly access the nematic order parameter. For example they cannot easily distinguish nematic order from the simultaneous stripe-like magnetic order found in many Fe SC. Because it is symmetry selective thanks to specific selection rules, Raman scattering has been shown to provide important insights into nematic degrees of freedom in Fe SC in both their normal and superconducting states even under nominally zero strain \cite{gallais_observation_2013,gallais_charge_2016,thorsmolle_critical_2016,kretzschmar_critical_2016,massat_charge-induced_2016}. However combining tunable anisotropic strain with symmetry resolved Raman scattering has so far been limited to measurements under fixed uniaxial stress, aimed mostly at obtaining detwinned crystals \cite{ren_nematic_2015,baum_interplay_2018}. Extending Raman measurements in Fe SC in a tunable uniaxial strain environment appears therefore desirable.

Here using a low temperature Raman scattering set-up under continuously variable uniaxial strain, hereafter called elasto-Raman scattering, we study the effect of uniaxial strain on nematic order by looking at Raman-active optical phonons. We focus on BaFe$_2$As$_2$, a prototypical Fe SC showing nematic order. We show that the As optical phonon can serve as a sensitive probe of the nematic order parameter under varying strain. First, using the anisotropy of the As phonon Raman tensor we demonstrate the monitoring of strain-induced orientation of nematic domains at low temperature. Second, above the orthorhombic-nematic transition temperature $T_{S}$, we show that the As phonon intensity can be used to track the strain-induced nematic order parameter, and demonstrate that it is a sensitive probe of the underlying electronic nematic susceptibility. Finally, we show that the nematic order parameter under strong strain follows the behavior expected for an Ising order parameter under a symmetry breaking. This indicates that the As phonon Raman intensity is mostly sensitive to nematic order as opposed to the stripe-like magnetic order. Complications in the quantitative interpretation of the data close to $T_{S}$, arising from the temperature and strain dependence of the elastic constants are also discussed. Our work establishes elasto-Raman scattering as a promising tool for anisotropic strain-control studies of the complex phase diagrams found in many quantum materials.

\section{Piezo-based elasto-Raman scattering set-up}

Dating back from the seminal work on electric field and anisotropic stress effects on Raman scattering phonon in semiconductors and insulators, there is a relatively long history of morphic effects, i.e. the reduction of the crystal symmetry by an applied force, in crystals \cite{fleury_electric-field-induced_1968,anastassakis_morphic_1971,ganesan_lattice_1970,melo_changes_1982,merle_uniaxial-stress_1980,anastassakis_effect_1993}. Early Raman studies under stress, and some recent ones on Fe SC \cite{ren_nematic_2015,baum_interplay_2018}, have been performed under constant uniaxial stress with limited tunability and control over the actual strain applied.
Our experimental set-up is based on an uniaxial piezoelectric cell (CS130 from Razorbill Instruments) capable of applying in-situ both tensile and compressive strains along a given direction \cite{hicks_piezoelectric-based_2014}. The BaFe$_2$As$_2$ single crystal (Ba122) (dimensions 1.6mm x 700 $\upmu$m x 80 $\upmu$m) is anchored at both ends using Loctite Stycast 2850FT epoxy (used with catalyst 24LV), each end being sandwiched in between two mounting plates. The strain cell is equipped with a capacitor based sensor which can monitor in-situ the change $\delta L$ in the gap between the two moving sample plates, giving access to the nominal strain applied to the crystal along the $x$ direction which we define as $\epsilon^{nom}_{xx}$=$\frac{\delta L}{L_0}$ where $L_0$ is the strain-free suspended sample length across the gap in the $x$ direction (Fig.~\ref{fig1}). The maximum amplitude variation of the gap is 10~$\upmu$m at 4~K, corresponding to a maximum nominal strain of $\epsilon^{nom}_{xx} \sim 10^{-2}$ for the present crystal. The configuration of the piezostacks is such that their thermal contraction is compensated thus minimizing the effect of temperature on the applied strain. Complications may arise from the differential thermal contraction between the titanium (Ti) body of the cell and the sample. For Ba122 however, this effect is minimal given the very similar temperature dependence of the expansion coefficient of Ti and Ba122 \cite{ikeda_symmetric_2018}. 
%Note that contrary to mechanical stress cells , it is the uniaxial strain that is controlled in our set-up. The distinction between constant stress and constant strain measurements is important for materials where elastic coefficient is strongly temperature dependent as in the case of Ba122 near $T_{S}$.
\par
The uniaxial strain cell was mounted on a specially design copper mount inside the vacuum of a close-cycled optical cryostat. To compensate the relatively low thermal conductivity of Ti, copper braids were used to thermally anchor the cell with the second stage of the cryocooler. The temperature was monitored using a silicon diode affixed directly on the cell body, next to the sample. Sample temperature as low as 6.5~K could be reached. The piezostacks voltages were controlled in-situ via a dual 200~V power supply, and the capacitance of the displacement sensor was monitored using a Keysight LCR meter (model E4980AL). The Raman scattering geometry was non-colinear, with the incoming photon wavevector arriving at 45\degree ~with respect to the normal of the sample surface plane, and the outgoing photon wavevector lying along along the normal (Fig.~\ref{fig1}(a)). A single longitudinal mode solid state laser emitting at 532~nm was used for the excitation. The laser spot size is estimated to be less than 50~$\upmu$m, so that we expect strain inhomogeneity to have less impact in our Raman measurements than in transport measurements performed in similar cells \cite{hicks_strong_2014,steppke_strong_2017,ikeda_symmetric_2018,malinowski_suppression_2020,worasaran_nematic_2021}. The outgoing photons were collected using a X20 long working distance microscope objective (NA=0.28) and analyzed via a triple grating spectrometer equipped with nitrogen cooled CCD camera. A laser power of 10 mW was used for all spectra. Based on previous Raman measurements on Ba122, the laser heating was estimated to be about 1~K/mW \cite{chauviere_raman_2011,kretzschmar_critical_2016}. The temperatures indicated in the following were corrected for this heating.

\section{Sample and Strain configuration}

\begin{figure}
    \centering
    \includegraphics[width=0.49\textwidth]{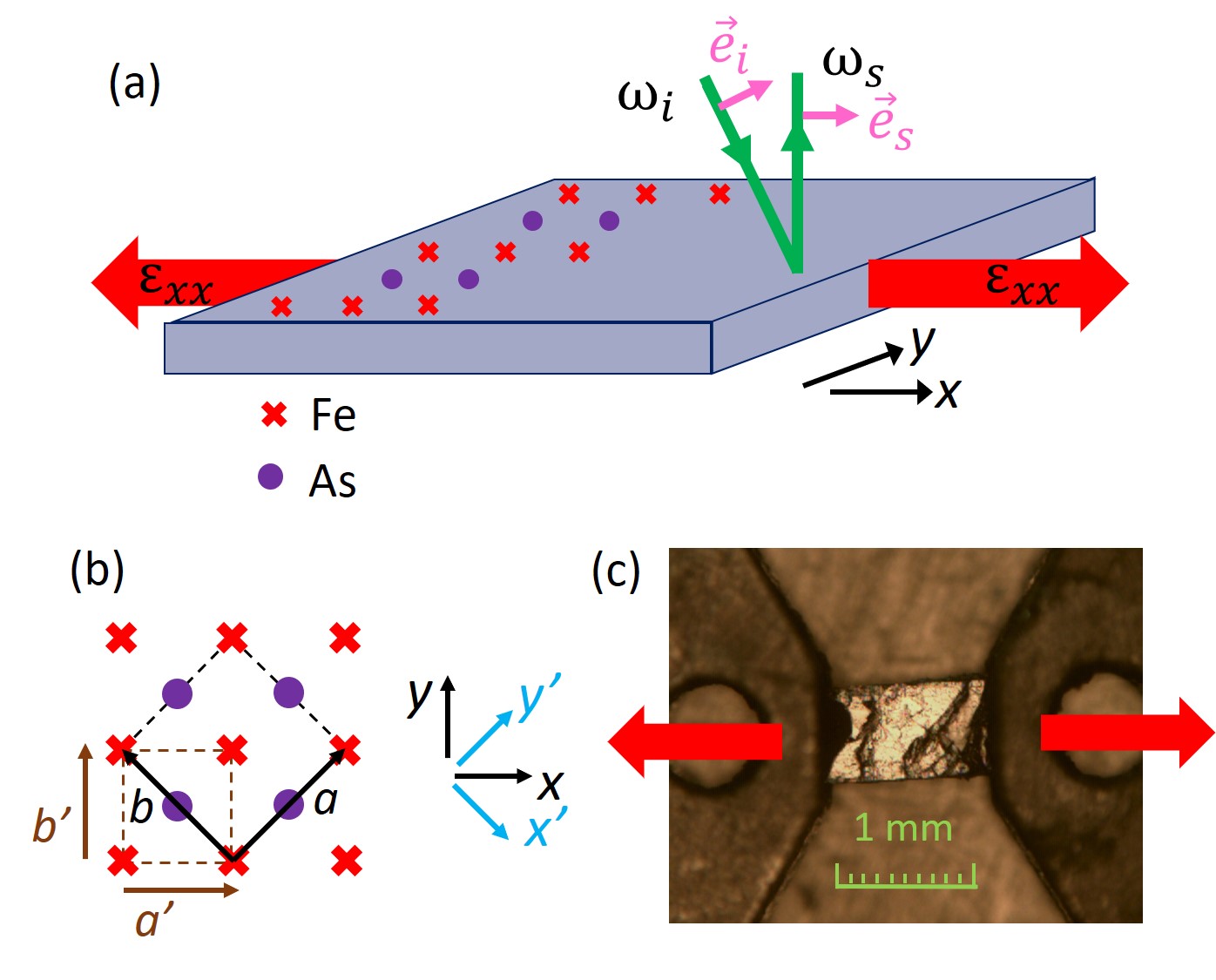}
    \caption{(a) Sketch of the sample and strain configuration. The green arrows denote the incident and scattered light propagation directions, the pink arrows the polarization directions. (b) Definition of the crystallographic axes $a$, $b$ (2 Fe tetragonal unit cell) and $a^{\prime}$, $b^{\prime}$ (1 Fe unit cell) with respect to the laboratory frames ($xy$) and ($x^{\prime}y^{\prime}$) (the latter being rotated by 45 \degree~with respect to $xy$). (c) Top view picture of the mounted sample. In (a) and (c), the red arrows denote the directions of the applied stress and the $\epsilon_{xx}$ strain.}
    \label{fig1}
\end{figure}

The Ba122 single crystal was cut-along the cristallographic [110] direction (along the Fe-Fe bonds) using a wire-saw. Transport measurements on single-crystals from the same batch indicate an orthorhombic-nematic transition (quasi-concomitant with a spin-density-wave (SDW) transition) at $T_S=138$~K. We define the direction of the applied strain $\epsilon_{xx}$ using the $xyz$ laboratory frame. In the 2 Fe unit cell of the FeAs plane, the tetragonal axes $a$ and $b$ are along the Fe-As bonds (Fig.~\ref{fig1}(b)). The strain $\epsilon_{xx}$ is applied at 45\degree~of the tetragonal axes and is therefore parallel to the directions of the axes $a'$ and $b'$ of the 1 Fe cell of the FeAs plane (Fig.~\ref{fig1}(b)). $\epsilon_{xx}$ directly couples to the nematic order parameter $\phi_{nem}$ of Ba122 which belongs to the $B_{2g}$ representation in the 2 Fe cell notation ($B_{1g}$ in the 1 Fe unit cell notation).
\par
Despite working at constant strain $\epsilon_{xx}$ and under uniaxial stress condition, we emphasize that, contrary to the stress, the actual strain applied is tri-axial. This is due to the finite Poisson ratio linking the strain along the applied direction $\epsilon_{xx}$ and the one along the orthogonal directions $\epsilon_{yy}$ and $\epsilon_{zz}$. The in-plane Poisson ratio is given by $\nu=-\frac{\epsilon_{yy}}{\epsilon_{xx}}$. An applied strain $\epsilon{}_{xx}$ along the [110] direction has both a nematic component $B_{2g}$ and an in-plane isotropic component $A_{1g}$: $\epsilon_{xx}=\epsilon_{A_{1g}}+\epsilon_{B_{2g}}$. The relationship between the applied strain and the $B_{2g}$ strain is given by $\epsilon_{B_{2g}}=\frac{1}{2}(1+\nu)\epsilon_{xx}$. A further complication is the strain transmission to the sample which is in general not perfect as a result of the strain within the epoxy glue \cite{hicks_piezoelectric-based_2014,ikeda_symmetric_2018,bartlett_relationship_2021}. As a consequence the actual strain experienced by the sample $\epsilon_{xx}$ will be in general different than the one measured by the capacitive sensor $\epsilon^{nom}_{xx}$ (see Appendix~\ref{Finite element simulation} for a discussion of these effects): $\epsilon_{xx}=\mu \epsilon_{xx}^{nom}$ where $\mu$ is the transmission ratio. Throughout the paper the data will be shown as a function of $\epsilon^{nom}_{xx}$, but the effects of both strain transmission ratio $\mu$ and Poisson coefficient $\nu$ are important and will be discussed when interpreting the data. Note however that, because of the similar thermal contraction between Ti and Ba122, at $\epsilon^{nom}_{xx} = 0$ we also have $\epsilon_{xx} = \epsilon_{B_{2g}} = 0$.
\par

\section{Arsenic optical Phonon and Raman selection rules}

In this paper we focus on the behavior of the Raman active optical phonon which involves the $c$-axis motion of the As atoms. The impact of strain on the electronic continuum will be discussed elsewhere. The strong sensitivity of the electronic properties of Fe SC to the As height is well-documented \cite{kuroki_pnictogen_2009,yildirim_strong_2009,yndurain_anomalous_2009,lee_effect_2008}. This is reflected in the significant impact of the nematic/SDW transition on this phonon lineshape and its Raman activity as demonstrated in previous Raman studies on pnictides and chalcogenides Fe SC \cite{chauviere_raman_2011,baum_interplay_2018,wu_-plane_2020,garcia-martinez_coupling_2013}. Here we will mainly be interested in the Raman selection rules of this phonon as a marker of the nematic order parameter. This will illustrate the interest of combining Raman selection rules with a tunable selective symmetry breaking field like anisotropic strain. 

Before proceeding with the results, we first briefly review the Raman selection rules for the As phonon in the tetragonal and orthorhombic/nematic phases. In the tetragonal phase the As phonon belongs to the $A_{1g}$ representation of the $D_{4h}$ point group with the corresponding Raman tensor (written in the tetragonal 2 Fe unit cell basis):
\begin{equation}
\Gamma^T_{As}=\begin{pmatrix}
\alpha^T_{aa} & 0 & 0\\
0 & \alpha^T_{aa} & 0 \\
0 & 0 & \alpha_{cc}
\end{pmatrix}
\end{equation}
where the first two diagonal components are identical because of the $C_4$ symmetry of the tetragonal lattice.
The components of the Raman tensor are given by the derivative of the dielectric tensor components $\hat{\epsilon}_{ij}$ (here we use the hat symbol to distinguish it from the strain) at the incoming photon frequency $\omega_L$ with respect to the phonon coordinates: $\alpha^T_{ij}=\frac{\partial \hat{\epsilon}_{ij}(\omega_L)}{\partial Q_{As}}$. Physically the As atomic motion modulates the electronic band structure and thus the dielectric constant at the incoming photon energy ($\sim$ 2.4 eV). This electronic mediated process makes Raman phonon intensities an indirect but very sensitive probe of the underlying electronic structure as in the case of As motion for Ba122 \cite{garcia-martinez_coupling_2013,baum_interplay_2018}. The Raman intensity for a given set of incoming ($\bm{e_i}$) and outgoing ($\bm{e_s}$) photon polarization vectors is proportional to the square of the tensor contracted by the photon polarization vectors:
\begin{equation}
I^{\bm{e_i}\bm{e_s}}_{As}=\lvert {\bm{e_i^*}}\Gamma_{As} {\bm{e_s}}\rvert^2    
\end{equation}
For $\bm{e_i}$ and $\bm{e_s}$ parallel along any cristallographic direction, we have $I^{\parallel}_{As}=\lvert \alpha^T_{aa} \rvert^2$, while for perpendicular polarizations $I^{\perp}_{As}=0$.
\par
In the nematic phase the As phonon belongs to the $A_{g}$ representation of the $D_{2h}$ point group. In this phase we use the usual orthorhombic axes along the Fe-Fe bonds $a'$ and $b'$ which are rotated by 45\degree ~with respect to the tetragonal axes (Fig.~\ref{fig1}(b)). These two directions are now inequivalent and the Raman tensor reads in the $a'b'$ coordinates (henceforth we will adopt the convention $a' < b'$ for the in-plane orthorhombic axes):
\begin{equation}
\Gamma^O_{As}=\begin{pmatrix}
\alpha^O_{a'a'} & 0 & 0\\
0 & \alpha^O_{b'b'} & 0 \\
0 & 0 & \alpha_{cc}
\end{pmatrix}
\end{equation}
where the in-plane diagonal components are now distinct because of $C_4$ symmetry breaking. Below $T_{S}$ the intensity for parallel polarization is no longer isotropic when using parallel in-plane polarizations and we have $I^{a'a'}_{As}$- $I^{b'b'}_{As}= \lvert \alpha^O_{a'a'} \rvert^2 - \lvert \alpha^O_{b'b'} \rvert^2\neq 0$.
Previous Raman studies of Ba122 have shown that the difference between $\alpha^O_{a'a'}$ and $\alpha^O_{b'b'}$ is extremely large below $T_{S}$\cite{chauviere_raman_2011,baum_interplay_2018,wu_-plane_2020}. Ab initio calculations indicate that electronic or magnetic anisotropy rather than the lattice distortion itself is responsible for the large difference in the Raman tensor components \cite{baum_interplay_2018}. The Raman tensor components can thus serve as local proxies of the electronic nematic order parameter as seen via the As atom displacement $\phi^{As}_{nem}$, defined as:
\begin{equation}
\phi^{As}_{nem}=\frac{\alpha^O_{a'a'} - \alpha^O_{b'b'}}{\alpha^O_{a'a'} + \alpha^O_{b'b'}}
\end{equation}
which has the same $B_{2g}$ symmetry as the electronic nematic order parameter. This quantity can be accessed from the measured Raman spectra in two different polarization configurations with incoming and outgoing photons respectively parallel to $a'$ and $b'$ orthorhombic axes:
\begin{equation}
I^{a'a'}_{As}- I^{b'b'}_{As}=\eta^2 \phi_{nem}^{As}
\end{equation}
where $\eta=\alpha^O_{a'a'} + \alpha^O_{b'b'}$. We note that $I^{a'a'}_{As}-I^{b'b'}_{As}$ will be sensitive to domain formation and will reflect the nematic order parameter $\phi_{nem}^{As}$ only in presence of a single nematic domain under the laser spot. 
\par

An alternative marker of the nematic order parameter is the activation of the As phonon intensity in crossed polarizations along the tetragonal axes $a$ and $b$: $I^{ab}_{As}=\lvert \alpha^O_{a'a'}-\alpha^O_{b'b'}\rvert^2\neq 0$.  $I^{ab}_{As}$ transforms like the square of the nematic order parameter: $\lvert \alpha^O_{a'a'}-\alpha^O_{b'b'}\rvert^2 \sim (\phi^{As}_{nem})^2$.  Contrary to the previous quantity, $I^{ab}_{As}$ is insensitive to domain formation and directly reflects $(\phi_{nem}^{As})^2$:
\begin{equation}
  I^{ab}_{As}=\eta^2 (\phi_{nem}^{As})^2  
\end{equation}

\par
In this work we will use $I^{a'a'}_{As}- I^{b'b'}_{As}$ to assess nematic domain population and $I^{ab}_{As}$ as a local marker of the nematic order parameter. Note that the directions of the orthorhombic axes $a'$ and $b'$ will in general be spatially dependent below $T_{S}$ due to domain formation. They will also depend on the applied strain. To avoid ambiguities we will therefore label the direction of photon polarizations using the laboratory frames $xyz$ and $x'y'z$ rather than using cristallographic axes. 

\section{Orientation of nematic twin-domains below $T_{S}$}

\begin{figure}
    \centering
    \includegraphics[width=0.49\textwidth]{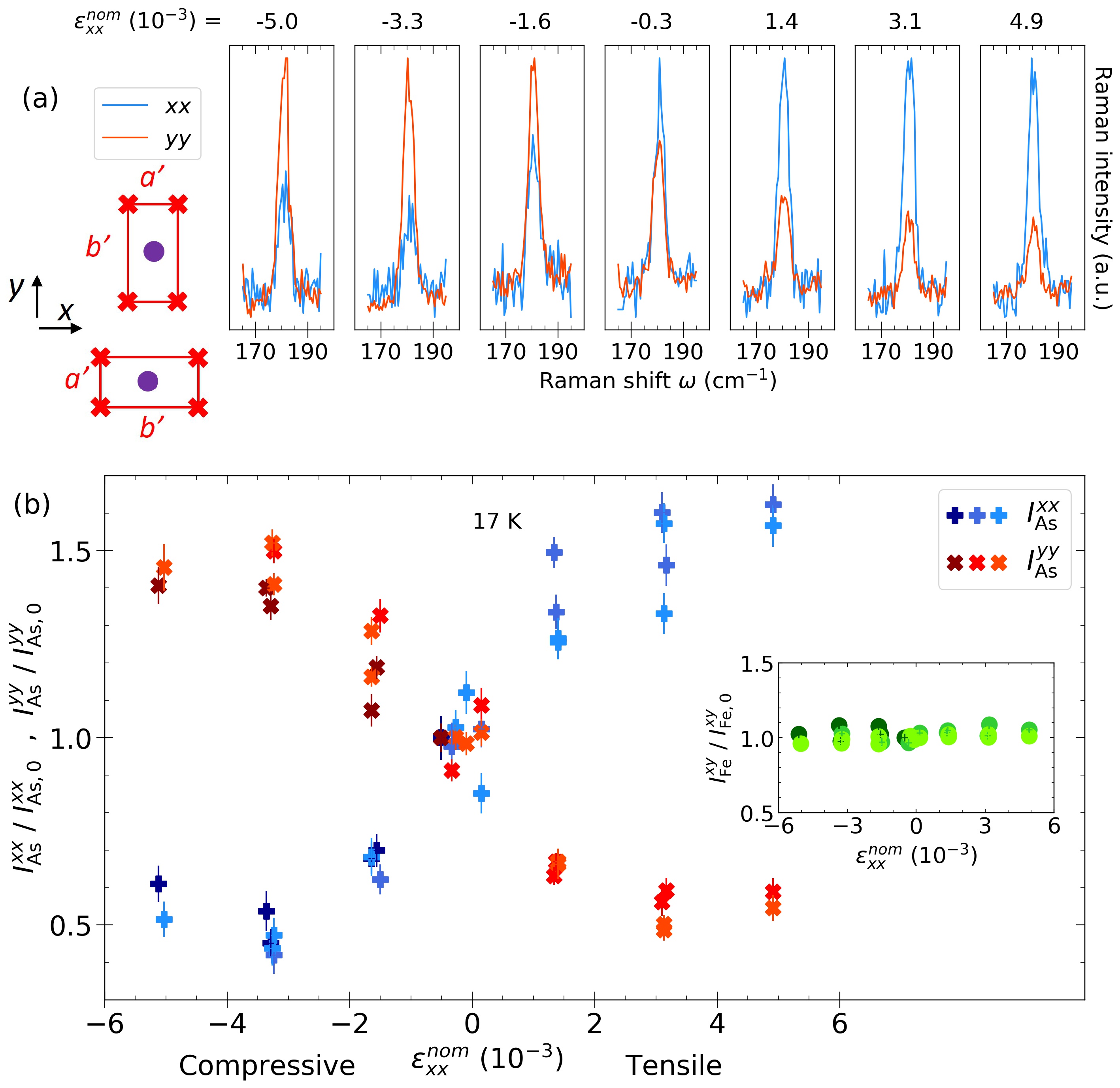}
    \caption{Orientation of nematic twin-domains at 17~K. (a) As phonon Raman spectra in the $xx$ and $yy$ polarization configurations for different strains. Here, the notations used for the polarization configurations indicate the direction of the incoming photon polarization followed by the direction of the outgoing photon polarization. The bottom left sketch depicts orientations of the two domains in the $xy$ laboratory frame. (b) Dependence of the As phonon intensities in $xx$ and $yy$ polarization configurations with respect to strain. Inset: dependence of the Fe phonon intensity in the $B_{1g}$ $xy$ polarization configuration with strain. We evaluated $I^{xx}_{As,0}$ as the mean of $I^{xx}_{As}$ data points for the strains close to $\epsilon^{nom}_{xx}=0$ (and likewise for $I^{yy}_{As,0}$ and $I^{xy}_{Fe,0}$).}
    \label{fig2}
\end{figure}

We first discuss the monitoring of nematic twin-domain orientations under strain. For this we use the As phonon polarization-resolved intensity at 17~K, deep in the nematic-orthorhombic phase. Fig. \ref{fig2}-(a) displays As phonon Raman spectra for different values of $\epsilon^{nom}_{xx}$ and for parallel polarizations along ($xx$) and perpendicular ($yy$) to the applied strain. The respective intensities $I^{xx}_{As}$ and $I^{yy}_{As}$ are strongly sensitive to the applied strain: for compressive strains $\epsilon^{nom}_{xx}<0$ we have $I^{xx}_{As} < I^{yy}_{As}$ but for tensile strain $I^{xx}_{As} > I^{yy}_{As}$. We define $I^{xx}_{As,0} = I^{xx}_{As}(\epsilon_{B_{2g}} = 0)$, and respectively for $I^{yy}_{As,0}$. The crossover is continuous with one polarization configuration gaining intensity at the expense of the other upon varying strain. By contrast the intensity of the $B_{1g}$ Fe optical phonon mode hardly changes with strain (see inset of Fig.~\ref{fig2}(b)).

\par
A quantitative evaluation of the As phonon intensities $I^{xx}_{As}$ and $I^{yy}_{As}$ as a function of strain can be obtained by fitting it with a Gaussian lineshape. The result is shown in Fig. \ref{fig2}(b) for three different cycles of compression and tension. The overall behavior is symmetric with respect to both strain and polarization directions and can be understood by a gradual modification of the population of nematic twin-domains under compression/tensile strain with the presence of two different types of domains, with the shorter axis $a'$ being along the laboratory frame direction $x$ or $y$ (Fig.~\ref{fig2}(a)). Calling $\beta(\epsilon^{nom}_{xx})$ the fraction of domains with the shorter axis $a'$ aligned along the direction of applied stress $x$, we have:
\begin{equation}
 I^{xx}_{As}=\beta \lvert \alpha^0_{a'a'} \rvert ^2+(1-\beta)\lvert \alpha^0_{b'b'} \rvert ^2
\end{equation}
 and equivalently: 
 \begin{equation}
   I^{yy}_{As}=\beta \lvert \alpha^0_{b'b'} \rvert ^2+(1-\beta)\lvert \alpha^0_{a'a'} \rvert ^2 
 \end{equation}
At zero strain the nematic domains are equiprobable giving $\beta(\epsilon^{nom}_{xx}=0)=0.5$ and therefore the phonon intensity is independent of the photon polarization $I^{xx}_{As,0}=I^{yy}_{As,0}=\frac{1}{2}(\lvert \alpha^0_{a'a'} \rvert ^2+\lvert \alpha^0_{b'b'} \rvert ^2)$ as observed experimentally. Under large strain the area under the laser spot becomes essentially single domain: the shorter orthorhombic axis $a'$ is parallel (perpendicular) to the $x$ direction for compression (tension). This results in $\beta=1$ (compression) or $\beta=0$ (tension). This yields $I^{xx}_{As}\propto\lvert\alpha^0_{a'a'}\rvert^{2}$ and $I^{yy}_{As}\propto\lvert\alpha^0_{b'b'}\rvert^{2}$ for strong compression, and  $I^{xx}_{As}\propto\lvert\alpha^0_{b'b'}\rvert^{2}$ and $I^{yy}_{As}\propto\lvert\alpha^0_{a'a'}\rvert^{2}$ for strong tension. Our data show that the modulus of the As Raman tensor element is larger along the longer orthorhombic axis:  $\lvert \alpha^0_{b'b'}\rvert ^2 \sim 3. \lvert \alpha^0_{a'a'} \rvert ^2$ in agreement with the Raman data under fixed compressive stress of Baum et al. \cite{baum_interplay_2018}. The overall evolution of the intensity can be rationalized by noting that we expect the domain fraction $\beta$ to follow a linear dependence as a function of strain: $\beta=\frac{1}{2}(1-\frac{\epsilon_{B_{2g}}}{\epsilon_s})$ where $\epsilon_s = \frac{a^{\prime} - b^{\prime}}{a^{\prime} + b^{\prime}}$ is the local orthorhombic strain within each domain. This gives a linear dependence of the phonon intensity with applied strain: 
\begin{equation}
I^{xx}_{As}=I^{xx}_{As,0} + \frac{\epsilon_{B_{2g}}}{2\epsilon_s}\eta^{2}\phi^{As}_{nem}
\end{equation}
The above relation is valid only for $\epsilon_{B_{2g}}<\epsilon_s$ in the twinned domain regime. In this regime the macroscopic strain of the crystal induced by the applied strain mostly occurs via domain wall motion with no changes in the lattice constants $a'$ and $b'$ nor in the nematic order parameter (i.e. $\phi^{As}_{nem}$ is constant). In this regime, the energy barrier for domain formation is probably low~\cite{bartlett_relationship_2021}, and indeed no significant hysteresis is observed in our data. We therefore expect the applied strain to be fully transmitted and entirely of $B_{2g}$ symmetry: $\epsilon^{nom}_{xx} \sim \epsilon_{xx} \sim \epsilon_{B_{2g}}$. Upon applying further strain however, we enter a new regime where macroscopic strain can only occur via a change in the orthorhombic distortion itself and its associated nematic order parameter $\phi^{As}_{nem}$.
\par
The transition between the two regimes is apparent in our data of Fig.~\ref{fig2}.
For large compressive and tensile strains (above $\sim \mp 3.10^{-3}$) the As phonon intensity becomes much more weakly dependent on the applied strain, indicating that the sample is close to being single domain. This value is close to the low temperature orthorhombic distortion $\epsilon_s$ of Ba122 observed by X-ray diffraction measurements \cite{rotter_spin-density-wave_2008}, confirming that the applied strain is fully transmitted to the crystal as a $B_{2g}$ strain in the low strain regime. When $\epsilon^{nom}_{xx}>\epsilon_s$, $\beta$ is fixed to 1 and 0 and the phonon intensity rather reflects the change of the nematic order parameter as probed by the local As Raman phonon tensor anisotropy $\phi^{As}_{nem}$ under applied strain, i.e. a form of local nematic susceptibility akin to the elastoresistivity measured by transport \cite{chu_divergent_2012}. This nematic susceptibility is expected to be rather small deep in the nematic phase where $\phi^{As}_{nem}$ is fully saturated \cite{chu_divergent_2012,gallais_observation_2013}. This likely explains the rather weak strain dependence of the phonon intensity above $\epsilon_s$. We will discuss this effect in the next section when presenting the data close to $T_S$.

\section{Strain-tuned nematicity around $T_{S}$}

\begin{figure*}
    \centering
    \includegraphics[width=0.85\textwidth]{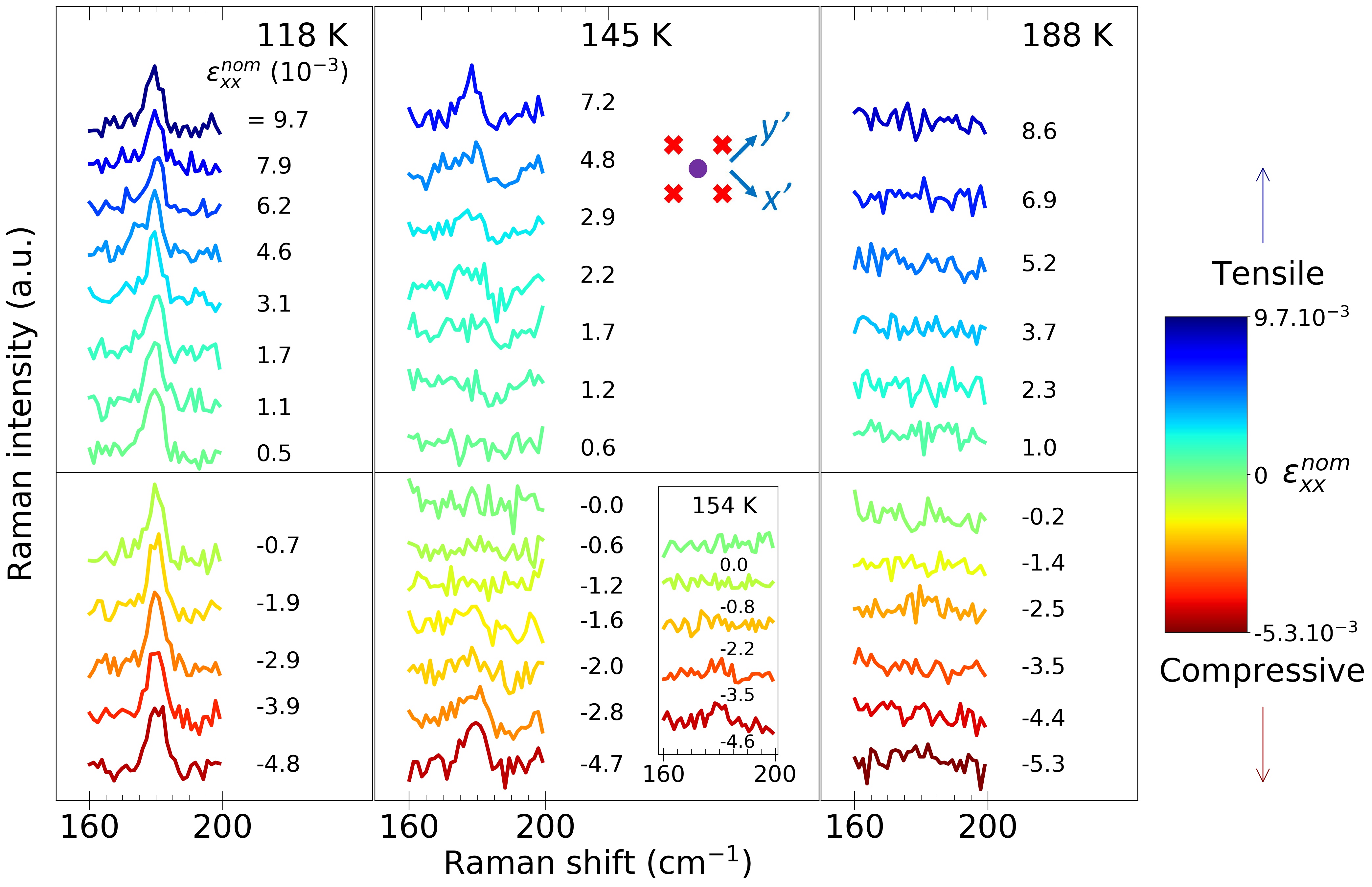}
    \caption{Strain dependence of the As phonon Raman spectra in the $x^{\prime}y^{\prime}$ polarization configuration at four temperatures around $T_{S}$. Note that crystal likely broke during the last measurement at 118~K under strong tension $\epsilon^{nom}_{xx}=9.7 \times 10^{-3}$. This precluded measurements at 154~K under tension: only spectra under compression are shown.}
    \label{fig3}
\end{figure*}

We now discuss the effect of strain on the As phonon close to the nematic transition temperature observed at $T_{S}=138$~K under nominally zero-strain. As phonon spectra at 118~K, 145~K, 154~K and 188~K in the $x^{\prime}y^{\prime}$ configuration, where incident and outgoing polarizations are perpendicular and along the $a$ and $b$ tetragonal axes, are shown as a function of strain in Fig. \ref{fig3}. The spectra at $T$=145~K $>T_S$ display the strongest strain dependence. At low strain no phonon is detected in this geometry as dictated by tetragonal symmetry: $I^{x^{\prime}y^{\prime}}_{As}=\lvert \alpha^O_{a'a'}-\alpha^O_{b'b'}\rvert^2 = 0$. Upon applying both compressive and tensile strain the phonon is activated in an approximate symmetric manner, indicating a strain-induced activation of the nematic order parameter $\phi^{As}_{nem}$ above $T_{S}$. Contrary to the spectra at 17~K discussed above, the activated phonon lineshape is asymmetric with a rather pronounced Fano lineshape that is evident at moderate strains, suggesting a significant coupling between the phonon and low energy electronic excitations. The Fano lineshape asymmetry was also reported close, but below $T_{S}$ in previous zero strain Raman data \cite{chauviere_raman_2011,wu_-plane_2020}. Increasing the temperature we observe a significant weakening of the strain induced activation of the As phonon. It is still visible at high strains at 154~K, but cannot be clearly resolved anymore within our signal-to-noise ratio at 188~K, except for $\epsilon_{xx}=-5.3 \times 10^{-3}$ where a weak and broad peak is detected. Note that the crystal broke during the measurements precluding measurements in tension at 154~K. Below $T_{S}$ at 118~K the phonon is visible at all strains and only weakly depends on strain.
\par
The evolution of the phonon intensity $I^{x^{\prime}y^{\prime}}_{As}$ as a function of strain and temperature
was evaluated by fitting the phonon using a Fano lineshape analysis. $\lvert \phi^{As}_{nem} \rvert \propto (I^{x^{\prime}y^{\prime}}_{As})^{\frac{1}{2}}$ is shown in Fig. \ref{fig4} for 118~K, 145~K and 154~K (see Appendix~\ref{Fano lineshape fitting} for details on the Fano lineshape fitting).

\begin{figure}
    \centering
    \includegraphics[width=0.49\textwidth]{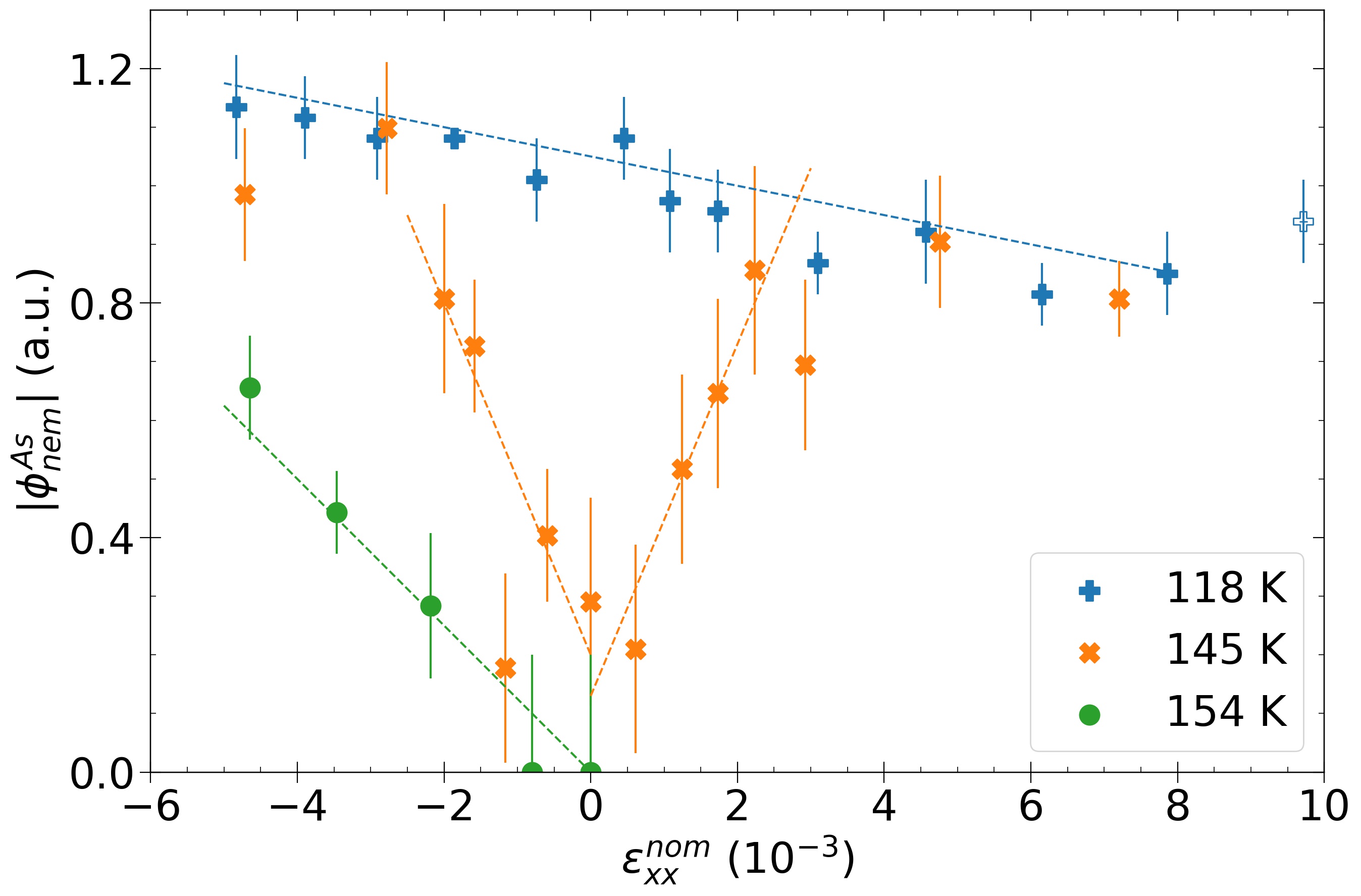}
    \caption{Strain dependence of the As phonon derived nematic order parameter $\lvert \phi^{As}_{nem} \rvert$ at three temperatures around $T_{S}$. The dashed lines are guides to the eye and not fits of the data. The empty symbol at 118~K at high tensile strain indicates that we suspect that the sample broke at this point.}
    \label{fig4}
\end{figure}

At 145~K $\lvert \phi^{As}_{nem} \rvert$ shows a steep and strain-symmetric activation around $\epsilon^{nom}_{xx}=0$. A much weaker effect is observed at 154~K with an initial slope $\frac{\partial \lvert \phi^{As}_{nem} \rvert}{\partial \epsilon_{xx}^{nom}}(\epsilon_{xx}^{nom}\rightarrow 0)$ at least 2 times smaller. The symmetric behavior observed at 145~K shows that the phonon intensity is mostly sensitive to the $B_{2g}$ component of the applied strain, in qualitative agreement with a strong nematic susceptibility close to $T_{S}$. At 118~K the intensity follows a much weaker strain-asymmetric dependence indicating a dominant effect coming from the isotropic $A_{1g}$ component of the applied strain. We assign this dependence to the monotonic decrease / increase of both $T_{S}$ and $T_N$ upon tensile / compressive $A_{1g}$ strain observed in transport measurements under in-plane uniaxial stress \cite{ikeda_symmetric_2018}. For strains above $\sim \pm 3 \times10^{-3}$ the phonon intensity at 145~K shows signs of saturation and reaches values close to the intensity observed below $T_{S}$ at 118~K. 
\par
At first sight and taking the initial slope $\frac{\partial \lvert \phi^{As}_{nem} \rvert}{\partial \epsilon_{xx}^{nom}}(\epsilon_{xx}^{nom}\rightarrow 0)$ as a marker of a nematic susceptibility $\chi_{nem}$, the results can be qualitatively understood in terms of a strongly temperature dependent electronic nematic susceptibility. The behavior of $\chi_{nem}$ above $T_{S}$ is well documented by previous elasto-resistivity, shear-modulus and zero-strain electronic Raman scattering measurements which all detect a clear enhancement of $\chi_{nem}$ close to $T_{S}$ in Ba122 \cite{chu_divergent_2012,yoshizawa_structural_2012,gallais_observation_2013}, in qualitative agreement with the observed behavior at 145, 154 and 188~K. At 118~K $\lvert \phi^{As}_{nem} \rvert$ does not show a symmetric behavior with respect $\epsilon_{xx}^{nom}$. Instead it displays a monotonic behavior, indicating a strongly reduced nematic susceptibility below $T_S$. Note that at 118~K the regime of domain orientation is likely very narrow, below $\lvert \epsilon^{nom}_{xx}\rvert= 2 \times 10^{-3}$, and has therefore a limited impact on the data presented in Fig. \ref{fig4}. Since transport measurements are plagued by domain formation below $T_{S}$, our data suggest that $\lvert \phi^{As}_{nem} \rvert$ is a potentially interesting alternative marker to track the nematic susceptibility and order parameter both above and below $T_{S}$. 
\par
There are however several caveats which must be addressed before linking the observed behavior to an enhanced nematic susceptibility. Formally a nematic susceptibility is defined as a derivative of a nematic observable with respect to a nematic field. In our case we can define it using our definition of the nematic order parameter as probed by the As phonon $\phi^{As}_{nem}$ as:
\begin{equation}
\chi^{As}_{nem}=\frac{\partial \phi_{nem}^{As}}{\partial \epsilon_{B_{2g}}}   
\end{equation}

The initial slope can be written in terms of the above-defined nematic susceptibility:
\begin{equation}
\frac{\partial \lvert \phi^{As}_{nem} \rvert}{\partial \epsilon_{B_{2g}}}=\lvert \eta \rvert \frac{\phi_{nem}^{As}}{\lvert \phi_{nem}^{As} \rvert}\chi_{nem}^{As}
\label{As-nem-suscept}
\end{equation}
The main complication when using equation~\eqref{As-nem-suscept} to access $\chi_{nem}^{As}$ from our data stems from converting the nominal strain measured $\epsilon_{xx}^{nom}$ into the actual $B_{2g}$ strain $\epsilon_{B_{2g}}$ experienced by the sample. Indeed we have:
\begin{equation}
 \epsilon_{B_{2g}}=\frac{1}{2}(1+\nu)\mu\epsilon^{nom}_{xx}  
\end{equation}
where $\mu$ is the transmission coefficient and $\nu$ the Poisson ratio. Since we are not interested in the absolute value of $\chi^{As}_{nem}$ but only in its strain and temperature dependence, in general it suffices that these coefficients are weakly dependent on both strain and temperature. In materials away from a structural instability this condition is satisfied, but this is clearly not the case near $T_{S}$ for Ba122. First the Poisson ratio $\nu$ is strongly temperature dependent near $T_{S}$: at $T_{S}$ the shear modulus $C_{66}\propto \frac{1-\nu}{1+\nu}$ goes to zero but it stiffens significantly away from $T_{S}$ \cite{yoshizawa_structural_2012}. As a consequence the applied strain is expected to be entirely $B_{2g}$ symmetric at $T_{S}$, but to quickly acquire a significant $A_{1g}$ component away from $T_{S}$. Fortunately in the low strain limit this effect can be taken into account via available data of the temperature dependence of elastic constant in Ba122 \cite{yoshizawa_anomalous_2012}. A second non-trivial effect comes from the transmission coefficient $\mu$. Because the Young modulus along the orthorhombic direction also softens to zero at $T_{S}$ \cite{bohmer_electronic_2016}, we expect the strain to be fully transmitted at $T_{S}$, but not away from it when the lattice hardens. Finite element simulations using available data on elastic constants were conducted in order to estimate the temperature dependence of $\mu$ (see Appendix~\ref{Finite element simulation}). 
%discuss FEM simulation and estimate combine effects true chi_nem at low strain. Is change of slope between 145K and 154K consistent with chi_nem from other sources ?
Overall the combination of these two effects gives corrections of $\sim1.3$ and $\sim 2$ at low strain between 145~K and 154~K, and 145~K and 188~K respectively. The data indicate at least a factor 2 between the slopes at 145~K and 154~K (Fig.~\ref{fig4}). Given the absence of any resolved As phonon at 188~K up to at least $\epsilon^{nom}_{xx} = 4 \times 10^{-3}$, we can give a lower bound estimate of at least a factor 4 between the experimental slopes at 145~K and 188~K. Therefore, we conclude that while these corrections are significant and must be accounted for in any quantitative discussion of the slope, they cannot fully account for the observed temperature dependence of the slopes at low strain. This observation confirms that $\lvert \phi^{As}_{nem} \rvert$ and $\chi^{As}_{nem}$ are not simple proxies of the lattice orthorhombicity $\epsilon_{B_{2g}}$ and shear modulus $C_{66}$ respectively, but have a significant component related to the purely electronic nematic order parameter and susceptibility.

\par
%discuss effects at high strain: stiffening of lattice / saturation
Beyond the small strain regime, the reasoning above cannot apply: because of the increase of the nematic order parameter, the shear modulus at high strain is expected to stiffen with respect to its low strain value. The effect is expected to be particularly strong near $T_{S}$ where the softening of the shear modulus will be significantly reduced at high strain and will recover its value at high temperature, far above the structural instability (see Appendix~\ref{Shear modulus under finite stress} for a discussion of these non-linear effects). Indeed recent elasto-X-ray scattering data in Co doped Ba122 indicate that both the transmission coefficient and the Poisson ratio become essentially temperature-independent for nominal strains above $\pm 3 \times 10^{-3}$~\cite{sanchez_spontaneous_2020}. In our case the strain induced reduction of the Poisson ratio and transmission coefficient at 145~K could partly explain the saturation of  $\lvert \phi^{As}_{nem} \rvert$ for $\lvert\epsilon_{xx}^{nom}\rvert> 4 \times10^{-3}$. However, disentangling this effect from the genuine behavior of $\chi_{nem}^{As}$ under strong applied strain is highly non-trivial.
\par
\begin{figure}
    \centering
    \includegraphics[width=0.47\textwidth]{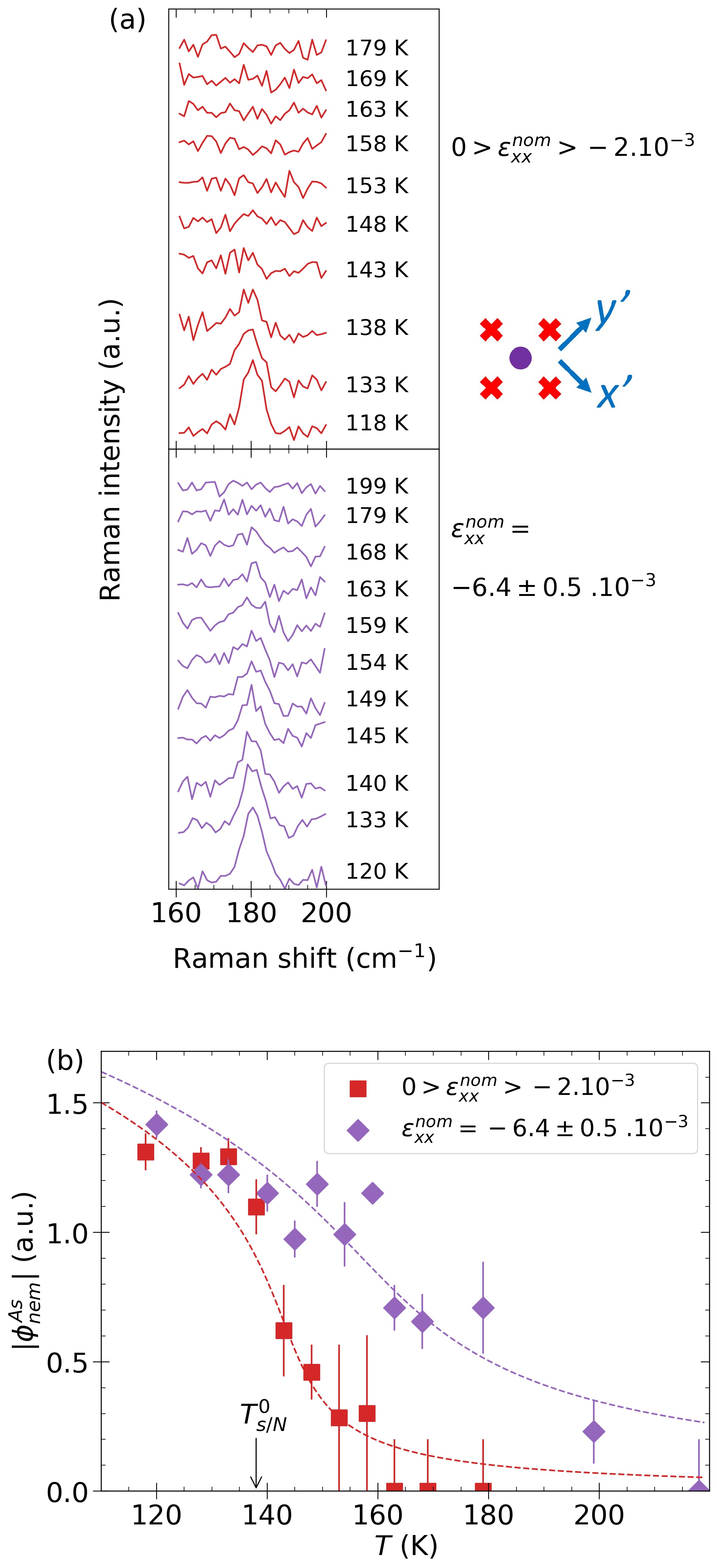}
    \caption{(a) Evolution of the As phonon Raman spectra in the $x^{\prime}y^{\prime}$ polarization configuration crossing $T_{S}$ at fixed strains. The evaluation of the strain for the small strain series (in red) is complicated by the fact that the sample was broken at that time. We estimated the strain value to be close to 0, with a maximum error evaluated by comparing the spectrum at 143~K with the series at 145~K displayed in Fig.~\ref{fig3}. (b) Dependence of As phonon nematic order parameter $\lvert \phi^{As}_{nem} \rvert$ with temperature at fixed strains. The dashed lines are guides to the eye following a Landau-type behavior of the order parameter with a symmetry-breaking field 5 times larger for the purple line compared to the red line (see Appendix~\ref{Nematic order parameter under a symmetry breaking field}).}    
    \label{fig5}
\end{figure}
%discuss issues with uniaxial stress measurement like neutron
An interesting corollary to the above discussion is the interpretation of temperature dependent measurements under constant uniaxial stress as reported in several neutron scattering and optical spectroscopy studies in various Fe SC materials \cite{dhital_effect_2012,lu_impact_2016,tam_uniaxial_2017,mirri_electrodynamic_2016,ren_nematic_2015,mirri_origin_2015}. Because of the strong softening of the Young's modulus along the orthorhombic axis close to $T_{S}$, the strain associated to the stress applied to the sample is strongly enhanced near $T_{S}$ and even diverges at $T_{S}$. This leads to much larger temperature dependent non-linear effects than by working under constant nominal strain, and makes the interpretation of constant stress measurements in terms of nematic susceptibilities problematic.
\par

%discuss T dep at high strain: analogy with Ising under field and origin of As anistropy: SDW versus nematic
Finally we discuss the origin of the strain induced activation of the As phonon in $x^{\prime}y^{\prime}$ geometry. Up to now we have assigned it to a finite nematic order parameter from symmetry based arguments. However the activation has in general been assigned to the emergence of the stripe-like SDW order at $T_N\approx T_S$, with only a minor role for the nematic order itself \cite{baum_interplay_2018,wu_-plane_2020}. In principle this question can be addressed in samples with sufficiently split $T_S$ and $T_N$ transitions, but unfortunately the As phonon intensity is in general too small in such samples to convincingly settle this issue. Here we address it by noting that the temperature dependence of the As phonon intensity under constant strain is expected to be rather different between the magnetic and nematic scenario. Indeed the nematic transition is no longer well-defined and becomes instead a crossover under $B_{2g}$ strain, as for a ferromagnet under an applied magnetic field. The SDW transition however, remains sharply defined since strain does not couple linearly to the magnetic order parameter.

 \par
The temperature dependence of $\lvert \phi^{As}_{nem} \rvert$ at constant strain and the corresponding spectra are displayed in Fig.~\ref{fig5}. Due to the crystal break before these temperature-dependent measurements, there is some uncertainty in the determination of the neutral point corresponding to $\epsilon_{xx}^{nom}=0$. Based on strain dependent spectra at fixed temperatures (Fig.~\ref{fig3}), we estimate the low strain measurements to be under small compressive strains, in absolute value smaller than 2.10$^{-3}$. As for the strain dependent measurements, the phonon was fitted using a Fano lineshape in order to extract its intensity (see Appendix~\ref{Fano lineshape fitting}). For small strains the phonon intensity has a relatively sharp onset close to the strain free $T_{S,N}= 138$~K, with a small tail above 140~K. Except for the small tail above 138~K, this agrees with previous Raman measurements under nominally zero strain \cite{chauviere_doping_2009,wu_coupling_2020}. At high compressive strain, $\epsilon^{nom}_{xx}\sim - 6.4 \times 10^{-3}$, the sharp onset is replaced by a very broad tail extending well above the strain free $T_{S,N}$. We note that Nuclear Magnetic Resonance (NMR) studies of Ba122 under uniaxial strain show a relatively weak effect of strain on $T_{N}$, with a shift of at most 10~K for $\epsilon^{nom}_{xx}\sim - 6 \times 10^{-3}$ \cite{kissikov_uniaxial_2018}. Within our experimental accuracy no additional jump or anomaly which could be assigned to $T_N$ is detected in the temperature dependence of $\lvert \phi^{As}_{nem} \rvert$. This indicates a minimal impact of the SDW order on the phonon intensity. Note however, that the effective phonon energy extracted from the Fano analysis bears fingerprints of $T_N$, as can be inferred from the change in phonon lineshape observed around 140~K (see Appendix~\ref{Fano lineshape fitting}). Overall the qualitative behavior of $\lvert \phi^{As}_{nem} \rvert$ is consistent with the temperature dependence of an Ising order parameter under a symmetry-breaking field, where the sharp phase transition at zero-field is replaced by a smooth crossover at high field (see lines in Fig.~\ref{fig5}(b) and Appendix~\ref{Nematic order parameter under a symmetry breaking field}). A potential complication in quantitatively analyzing the data is the fact that despite working at constant $\epsilon^{nom}_{xx}$, the $B_{2g}$ strain could be temperature dependent due to the softening of the lattice. This effect is certainly present at moderate strains. However, as we argue in Appendix~\ref{Shear modulus under finite stress} we expect the lattice to stiffen considerably at these high strain levels. In that case the shear modulus temperature dependence becomes much weaker \cite{bohmer_electronic_2016,sanchez_spontaneous_2020} and likely plays a marginal role in the observed temperature dependence for strain above 5 $\times 10^{-3}$. We therefore conclude that in Ba122 the anisotropy of the Raman tensor elements as measured via $\lvert \phi^{As}_{nem} \rvert$ reflects nematic rather than magnetic degrees of freedom. As such this quantity appears to be a faithful probe of the electronic nematic order parameter.
\par
%remaining questions: role of lattice versus electronic dof in \phi_nem^As (indirect), decoupling of shear modulus / chi_nem at high strain (Sanchez et al.)

In conclusion using a novel low temperature polarization-resolved elasto-Raman set-up we have shown that the As phonon can provide a valuable local probe of nematic domains and order parameter in the model system Ba122. This has allowed us to track the nematic order parameter both as a function of strain and temperature above and below the nematic transition temperature. We have also stressed some possible pitfalls that must be taken into account when interpreting uniaxial strain data near a structural instability. Our work illustrates the interest of combining a symmetry-resolved stimulus like uniaxial strain with a symmetry-resolved probe like Raman. While the As phonon remains a rather indirect way to probe electronic nematic degrees of freedom in a metallic system with respect to elasto-resisivity and electronic Raman scattering for e.g., our data demonstrate that elasto-Raman studies of optical phonon could be a valuable tool to study nematic degrees of freedom in insulators where the aforementioned techniques are not accessible.

\section{Ackowledgements}
The authors thank Indranil Paul for helpful discussions and a careful reading of the manuscript.
\bibliography{biblio}

\begin{thebibliography}{59}
\expandafter\ifx\csname natexlab\endcsname\relax\def\natexlab#1{#1}\fi
\expandafter\ifx\csname bibnamefont\endcsname\relax
  \def\bibnamefont#1{#1}\fi
\expandafter\ifx\csname bibfnamefont\endcsname\relax
  \def\bibfnamefont#1{#1}\fi
\expandafter\ifx\csname citenamefont\endcsname\relax
  \def\citenamefont#1{#1}\fi
\expandafter\ifx\csname url\endcsname\relax
  \def\url#1{\texttt{#1}}\fi
\expandafter\ifx\csname urlprefix\endcsname\relax\def\urlprefix{URL }\fi
\providecommand{\bibinfo}[2]{#2}
\providecommand{\eprint}[2][]{\url{#2}}

\bibitem[{\citenamefont{Chu et~al.}(2012)\citenamefont{Chu, Kuo, Analytis, and
  Fisher}}]{chu_divergent_2012}
\bibinfo{author}{\bibfnamefont{J.-H.} \bibnamefont{Chu}},
  \bibinfo{author}{\bibfnamefont{H.-H.} \bibnamefont{Kuo}},
  \bibinfo{author}{\bibfnamefont{J.~G.} \bibnamefont{Analytis}},
  \bibnamefont{and} \bibinfo{author}{\bibfnamefont{I.~R.}
  \bibnamefont{Fisher}}, \bibinfo{journal}{Science}
  \textbf{\bibinfo{volume}{337}}, \bibinfo{pages}{710} (\bibinfo{year}{2012}).

\bibitem[{\citenamefont{Hicks et~al.}(2014{\natexlab{a}})\citenamefont{Hicks,
  Brodsky, Yelland, Gibbs, Bruin, Barber, Edkins, Nishimura, Yonezawa, Maeno
  et~al.}}]{hicks_strong_2014}
\bibinfo{author}{\bibfnamefont{C.~W.} \bibnamefont{Hicks}},
  \bibinfo{author}{\bibfnamefont{D.~O.} \bibnamefont{Brodsky}},
  \bibinfo{author}{\bibfnamefont{E.~A.} \bibnamefont{Yelland}},
  \bibinfo{author}{\bibfnamefont{A.~S.} \bibnamefont{Gibbs}},
  \bibinfo{author}{\bibfnamefont{J.~A.~N.} \bibnamefont{Bruin}},
  \bibinfo{author}{\bibfnamefont{M.~E.} \bibnamefont{Barber}},
  \bibinfo{author}{\bibfnamefont{S.~D.} \bibnamefont{Edkins}},
  \bibinfo{author}{\bibfnamefont{K.}~\bibnamefont{Nishimura}},
  \bibinfo{author}{\bibfnamefont{S.}~\bibnamefont{Yonezawa}},
  \bibinfo{author}{\bibfnamefont{Y.}~\bibnamefont{Maeno}},
  \bibnamefont{et~al.}, \bibinfo{journal}{Science}
  \textbf{\bibinfo{volume}{344}}, \bibinfo{pages}{283}
  (\bibinfo{year}{2014}{\natexlab{a}}).

\bibitem[{\citenamefont{Steppke et~al.}(2017)\citenamefont{Steppke, Zhao,
  Barber, Scaffidi, Jerzembeck, Rosner, Gibbs, Maeno, Simon, Mackenzie
  et~al.}}]{steppke_strong_2017}
\bibinfo{author}{\bibfnamefont{A.}~\bibnamefont{Steppke}},
  \bibinfo{author}{\bibfnamefont{L.}~\bibnamefont{Zhao}},
  \bibinfo{author}{\bibfnamefont{M.~E.} \bibnamefont{Barber}},
  \bibinfo{author}{\bibfnamefont{T.}~\bibnamefont{Scaffidi}},
  \bibinfo{author}{\bibfnamefont{F.}~\bibnamefont{Jerzembeck}},
  \bibinfo{author}{\bibfnamefont{H.}~\bibnamefont{Rosner}},
  \bibinfo{author}{\bibfnamefont{A.~S.} \bibnamefont{Gibbs}},
  \bibinfo{author}{\bibfnamefont{Y.}~\bibnamefont{Maeno}},
  \bibinfo{author}{\bibfnamefont{S.~H.} \bibnamefont{Simon}},
  \bibinfo{author}{\bibfnamefont{A.~P.} \bibnamefont{Mackenzie}},
  \bibnamefont{et~al.}, \bibinfo{journal}{Science}
  \textbf{\bibinfo{volume}{355}}, \bibinfo{pages}{eaaf9398}
  (\bibinfo{year}{2017}).

\bibitem[{\citenamefont{Kim et~al.}(2018)\citenamefont{Kim, Souliou, Barber,
  Lefrançois, Minola, Tortora, Heid, Nandi, Borzi, Garbarino
  et~al.}}]{kim_uniaxial_2018}
\bibinfo{author}{\bibfnamefont{H.-H.} \bibnamefont{Kim}},
  \bibinfo{author}{\bibfnamefont{S.~M.} \bibnamefont{Souliou}},
  \bibinfo{author}{\bibfnamefont{M.~E.} \bibnamefont{Barber}},
  \bibinfo{author}{\bibfnamefont{E.}~\bibnamefont{Lefrançois}},
  \bibinfo{author}{\bibfnamefont{M.}~\bibnamefont{Minola}},
  \bibinfo{author}{\bibfnamefont{M.}~\bibnamefont{Tortora}},
  \bibinfo{author}{\bibfnamefont{R.}~\bibnamefont{Heid}},
  \bibinfo{author}{\bibfnamefont{N.}~\bibnamefont{Nandi}},
  \bibinfo{author}{\bibfnamefont{R.~A.} \bibnamefont{Borzi}},
  \bibinfo{author}{\bibfnamefont{G.}~\bibnamefont{Garbarino}},
  \bibnamefont{et~al.}, \bibinfo{journal}{Science}
  \textbf{\bibinfo{volume}{362}}, \bibinfo{pages}{1040} (\bibinfo{year}{2018}).

\bibitem[{\citenamefont{Fernandes et~al.}(2014)\citenamefont{Fernandes,
  Chubukov, and Schmalian}}]{fernandes_what_2014}
\bibinfo{author}{\bibfnamefont{R.~M.} \bibnamefont{Fernandes}},
  \bibinfo{author}{\bibfnamefont{A.~V.} \bibnamefont{Chubukov}},
  \bibnamefont{and}
  \bibinfo{author}{\bibfnamefont{J.}~\bibnamefont{Schmalian}},
  \bibinfo{journal}{Nature Physics} \textbf{\bibinfo{volume}{10}},
  \bibinfo{pages}{97} (\bibinfo{year}{2014}).

\bibitem[{\citenamefont{Hosoi et~al.}(2016)\citenamefont{Hosoi, Matsuura,
  Ishida, Wang, Mizukami, Watashige, Kasahara, Matsuda, and
  Shibauchi}}]{hosoi_nematic_2016}
\bibinfo{author}{\bibfnamefont{S.}~\bibnamefont{Hosoi}},
  \bibinfo{author}{\bibfnamefont{K.}~\bibnamefont{Matsuura}},
  \bibinfo{author}{\bibfnamefont{K.}~\bibnamefont{Ishida}},
  \bibinfo{author}{\bibfnamefont{H.}~\bibnamefont{Wang}},
  \bibinfo{author}{\bibfnamefont{Y.}~\bibnamefont{Mizukami}},
  \bibinfo{author}{\bibfnamefont{T.}~\bibnamefont{Watashige}},
  \bibinfo{author}{\bibfnamefont{S.}~\bibnamefont{Kasahara}},
  \bibinfo{author}{\bibfnamefont{Y.}~\bibnamefont{Matsuda}}, \bibnamefont{and}
  \bibinfo{author}{\bibfnamefont{T.}~\bibnamefont{Shibauchi}},
  \bibinfo{journal}{Proceedings of the National Academy of Sciences}
  \textbf{\bibinfo{volume}{113}}, \bibinfo{pages}{8139} (\bibinfo{year}{2016}).

\bibitem[{\citenamefont{Kuo et~al.}(2016)\citenamefont{Kuo, Chu, Palmstrom,
  Kivelson, and Fisher}}]{kuo_ubiquitous_2016}
\bibinfo{author}{\bibfnamefont{H.-H.} \bibnamefont{Kuo}},
  \bibinfo{author}{\bibfnamefont{J.-H.} \bibnamefont{Chu}},
  \bibinfo{author}{\bibfnamefont{J.~C.} \bibnamefont{Palmstrom}},
  \bibinfo{author}{\bibfnamefont{S.~A.} \bibnamefont{Kivelson}},
  \bibnamefont{and} \bibinfo{author}{\bibfnamefont{I.~R.}
  \bibnamefont{Fisher}}, \bibinfo{journal}{Science}
  \textbf{\bibinfo{volume}{352}}, \bibinfo{pages}{958} (\bibinfo{year}{2016}).

\bibitem[{\citenamefont{Malinowski et~al.}(2020)\citenamefont{Malinowski,
  Jiang, Sanchez, Mutch, Liu, Went, Liu, Ryan, Kim, and
  Chu}}]{malinowski_suppression_2020}
\bibinfo{author}{\bibfnamefont{P.}~\bibnamefont{Malinowski}},
  \bibinfo{author}{\bibfnamefont{Q.}~\bibnamefont{Jiang}},
  \bibinfo{author}{\bibfnamefont{J.~J.} \bibnamefont{Sanchez}},
  \bibinfo{author}{\bibfnamefont{J.}~\bibnamefont{Mutch}},
  \bibinfo{author}{\bibfnamefont{Z.}~\bibnamefont{Liu}},
  \bibinfo{author}{\bibfnamefont{P.}~\bibnamefont{Went}},
  \bibinfo{author}{\bibfnamefont{J.}~\bibnamefont{Liu}},
  \bibinfo{author}{\bibfnamefont{P.~J.} \bibnamefont{Ryan}},
  \bibinfo{author}{\bibfnamefont{J.-W.} \bibnamefont{Kim}}, \bibnamefont{and}
  \bibinfo{author}{\bibfnamefont{J.-H.} \bibnamefont{Chu}},
  \bibinfo{journal}{Nature Physics} p. \bibinfo{pages}{1189}
  (\bibinfo{year}{2020}).

\bibitem[{\citenamefont{Maharaj et~al.}(2017)\citenamefont{Maharaj, Rosenberg,
  Hristov, Berg, Fernandes, Fisher, and Kivelson}}]{maharaj_transverse_2017}
\bibinfo{author}{\bibfnamefont{A.~V.} \bibnamefont{Maharaj}},
  \bibinfo{author}{\bibfnamefont{E.~W.} \bibnamefont{Rosenberg}},
  \bibinfo{author}{\bibfnamefont{A.~T.} \bibnamefont{Hristov}},
  \bibinfo{author}{\bibfnamefont{E.}~\bibnamefont{Berg}},
  \bibinfo{author}{\bibfnamefont{R.~M.} \bibnamefont{Fernandes}},
  \bibinfo{author}{\bibfnamefont{I.~R.} \bibnamefont{Fisher}},
  \bibnamefont{and} \bibinfo{author}{\bibfnamefont{S.~A.}
  \bibnamefont{Kivelson}}, \bibinfo{journal}{Proceedings of the National
  Academy of Sciences} \textbf{\bibinfo{volume}{114}}, \bibinfo{pages}{13430}
  (\bibinfo{year}{2017}).

\bibitem[{\citenamefont{Worasaran et~al.}(2021)\citenamefont{Worasaran, Ikeda,
  Palmstrom, Straquadine, Kivelson, and Fisher}}]{worasaran_nematic_2021}
\bibinfo{author}{\bibfnamefont{T.}~\bibnamefont{Worasaran}},
  \bibinfo{author}{\bibfnamefont{M.~S.} \bibnamefont{Ikeda}},
  \bibinfo{author}{\bibfnamefont{J.~C.} \bibnamefont{Palmstrom}},
  \bibinfo{author}{\bibfnamefont{J.~A.~W.} \bibnamefont{Straquadine}},
  \bibinfo{author}{\bibfnamefont{S.~A.} \bibnamefont{Kivelson}},
  \bibnamefont{and} \bibinfo{author}{\bibfnamefont{I.~R.}
  \bibnamefont{Fisher}}, \bibinfo{journal}{Science}
  \textbf{\bibinfo{volume}{372}}, \bibinfo{pages}{973} (\bibinfo{year}{2021}).

\bibitem[{\citenamefont{Kuo et~al.}(2013)\citenamefont{Kuo, Shapiro, Riggs, and
  Fisher}}]{kuo_measurement_2013}
\bibinfo{author}{\bibfnamefont{H.-H.} \bibnamefont{Kuo}},
  \bibinfo{author}{\bibfnamefont{M.~C.} \bibnamefont{Shapiro}},
  \bibinfo{author}{\bibfnamefont{S.~C.} \bibnamefont{Riggs}}, \bibnamefont{and}
  \bibinfo{author}{\bibfnamefont{I.~R.} \bibnamefont{Fisher}},
  \bibinfo{journal}{Physical Review B} \textbf{\bibinfo{volume}{88}},
  \bibinfo{pages}{085113} (\bibinfo{year}{2013}).

\bibitem[{\citenamefont{Hicks et~al.}(2014{\natexlab{b}})\citenamefont{Hicks,
  Barber, Edkins, Brodsky, and Mackenzie}}]{hicks_piezoelectric-based_2014}
\bibinfo{author}{\bibfnamefont{C.~W.} \bibnamefont{Hicks}},
  \bibinfo{author}{\bibfnamefont{M.~E.} \bibnamefont{Barber}},
  \bibinfo{author}{\bibfnamefont{S.~D.} \bibnamefont{Edkins}},
  \bibinfo{author}{\bibfnamefont{D.~O.} \bibnamefont{Brodsky}},
  \bibnamefont{and} \bibinfo{author}{\bibfnamefont{A.~P.}
  \bibnamefont{Mackenzie}}, \bibinfo{journal}{Review of Scientific Instruments}
  \textbf{\bibinfo{volume}{85}}, \bibinfo{pages}{065003}
  (\bibinfo{year}{2014}{\natexlab{b}}).

\bibitem[{\citenamefont{Bartlett et~al.}(2021)\citenamefont{Bartlett, Steppke,
  Hosoi, Noad, Park, Timm, Shibauchi, Mackenzie, and
  Hicks}}]{bartlett_relationship_2021}
\bibinfo{author}{\bibfnamefont{J.~M.} \bibnamefont{Bartlett}},
  \bibinfo{author}{\bibfnamefont{A.}~\bibnamefont{Steppke}},
  \bibinfo{author}{\bibfnamefont{S.}~\bibnamefont{Hosoi}},
  \bibinfo{author}{\bibfnamefont{H.}~\bibnamefont{Noad}},
  \bibinfo{author}{\bibfnamefont{J.}~\bibnamefont{Park}},
  \bibinfo{author}{\bibfnamefont{C.}~\bibnamefont{Timm}},
  \bibinfo{author}{\bibfnamefont{T.}~\bibnamefont{Shibauchi}},
  \bibinfo{author}{\bibfnamefont{A.~P.} \bibnamefont{Mackenzie}},
  \bibnamefont{and} \bibinfo{author}{\bibfnamefont{C.~W.} \bibnamefont{Hicks}},
  \bibinfo{journal}{Physical Review X} \textbf{\bibinfo{volume}{11}},
  \bibinfo{pages}{021038} (\bibinfo{year}{2021}).

\bibitem[{\citenamefont{Mirri et~al.}(2015)\citenamefont{Mirri, Dusza,
  Bastelberger, Chinotti, Degiorgi, Chu, Kuo, and Fisher}}]{mirri_origin_2015}
\bibinfo{author}{\bibfnamefont{C.}~\bibnamefont{Mirri}},
  \bibinfo{author}{\bibfnamefont{A.}~\bibnamefont{Dusza}},
  \bibinfo{author}{\bibfnamefont{S.}~\bibnamefont{Bastelberger}},
  \bibinfo{author}{\bibfnamefont{M.}~\bibnamefont{Chinotti}},
  \bibinfo{author}{\bibfnamefont{L.}~\bibnamefont{Degiorgi}},
  \bibinfo{author}{\bibfnamefont{J.-H.} \bibnamefont{Chu}},
  \bibinfo{author}{\bibfnamefont{H.-H.} \bibnamefont{Kuo}}, \bibnamefont{and}
  \bibinfo{author}{\bibfnamefont{I.}~\bibnamefont{Fisher}},
  \bibinfo{journal}{Physical Review Letters} \textbf{\bibinfo{volume}{115}},
  \bibinfo{pages}{107001} (\bibinfo{year}{2015}).

\bibitem[{\citenamefont{Mirri et~al.}(2016)\citenamefont{Mirri, Dusza,
  Bastelberger, Chinotti, Chu, Kuo, Fisher, and
  Degiorgi}}]{mirri_electrodynamic_2016}
\bibinfo{author}{\bibfnamefont{C.}~\bibnamefont{Mirri}},
  \bibinfo{author}{\bibfnamefont{A.}~\bibnamefont{Dusza}},
  \bibinfo{author}{\bibfnamefont{S.}~\bibnamefont{Bastelberger}},
  \bibinfo{author}{\bibfnamefont{M.}~\bibnamefont{Chinotti}},
  \bibinfo{author}{\bibfnamefont{J.-H.} \bibnamefont{Chu}},
  \bibinfo{author}{\bibfnamefont{H.-H.} \bibnamefont{Kuo}},
  \bibinfo{author}{\bibfnamefont{I.~R.} \bibnamefont{Fisher}},
  \bibnamefont{and} \bibinfo{author}{\bibfnamefont{L.}~\bibnamefont{Degiorgi}},
  \bibinfo{journal}{Physical Review B} \textbf{\bibinfo{volume}{93}},
  \bibinfo{pages}{085114} (\bibinfo{year}{2016}).

\bibitem[{\citenamefont{Kissikov et~al.}(2017)\citenamefont{Kissikov, Sarkar,
  Bush, Lawson, Canfield, and Curro}}]{kissikov_nuclear_2017}
\bibinfo{author}{\bibfnamefont{T.}~\bibnamefont{Kissikov}},
  \bibinfo{author}{\bibfnamefont{R.}~\bibnamefont{Sarkar}},
  \bibinfo{author}{\bibfnamefont{B.~T.} \bibnamefont{Bush}},
  \bibinfo{author}{\bibfnamefont{M.}~\bibnamefont{Lawson}},
  \bibinfo{author}{\bibfnamefont{P.~C.} \bibnamefont{Canfield}},
  \bibnamefont{and} \bibinfo{author}{\bibfnamefont{N.~J.} \bibnamefont{Curro}},
  \bibinfo{journal}{Review of Scientific Instruments}
  \textbf{\bibinfo{volume}{88}}, \bibinfo{pages}{103902}
  (\bibinfo{year}{2017}).

\bibitem[{\citenamefont{Kissikov et~al.}(2018)\citenamefont{Kissikov, Sarkar,
  Lawson, Bush, Timmons, Tanatar, Prozorov, Bud’ko, Canfield, Fernandes
  et~al.}}]{kissikov_uniaxial_2018}
\bibinfo{author}{\bibfnamefont{T.}~\bibnamefont{Kissikov}},
  \bibinfo{author}{\bibfnamefont{R.}~\bibnamefont{Sarkar}},
  \bibinfo{author}{\bibfnamefont{M.}~\bibnamefont{Lawson}},
  \bibinfo{author}{\bibfnamefont{B.~T.} \bibnamefont{Bush}},
  \bibinfo{author}{\bibfnamefont{E.~I.} \bibnamefont{Timmons}},
  \bibinfo{author}{\bibfnamefont{M.~A.} \bibnamefont{Tanatar}},
  \bibinfo{author}{\bibfnamefont{R.}~\bibnamefont{Prozorov}},
  \bibinfo{author}{\bibfnamefont{S.~L.} \bibnamefont{Bud’ko}},
  \bibinfo{author}{\bibfnamefont{P.~C.} \bibnamefont{Canfield}},
  \bibinfo{author}{\bibfnamefont{R.~M.} \bibnamefont{Fernandes}},
  \bibnamefont{et~al.}, \bibinfo{journal}{Nature Communications}
  \textbf{\bibinfo{volume}{9}}, \bibinfo{pages}{1058} (\bibinfo{year}{2018}).

\bibitem[{\citenamefont{Andrade et~al.}(2018)\citenamefont{Andrade, Berger,
  Rosenthal, Wang, Xing, Wang, Jin, Fernandes, Millis, and
  Pasupathy}}]{andrade_visualizing_2018}
\bibinfo{author}{\bibfnamefont{E.~F.} \bibnamefont{Andrade}},
  \bibinfo{author}{\bibfnamefont{A.~N.} \bibnamefont{Berger}},
  \bibinfo{author}{\bibfnamefont{E.~P.} \bibnamefont{Rosenthal}},
  \bibinfo{author}{\bibfnamefont{X.}~\bibnamefont{Wang}},
  \bibinfo{author}{\bibfnamefont{L.}~\bibnamefont{Xing}},
  \bibinfo{author}{\bibfnamefont{X.}~\bibnamefont{Wang}},
  \bibinfo{author}{\bibfnamefont{C.}~\bibnamefont{Jin}},
  \bibinfo{author}{\bibfnamefont{R.~M.} \bibnamefont{Fernandes}},
  \bibinfo{author}{\bibfnamefont{A.~J.} \bibnamefont{Millis}},
  \bibnamefont{and} \bibinfo{author}{\bibfnamefont{A.~N.}
  \bibnamefont{Pasupathy}}, \bibinfo{journal}{arXiv:1812.05287 [cond-mat]}
  (\bibinfo{year}{2018}).

\bibitem[{\citenamefont{Pfau et~al.}(2019)\citenamefont{Pfau, Chen, Yi,
  Hashimoto, Rotundu, Palmstrom, Chen, Dai, Straquadine, Hristov
  et~al.}}]{pfau_momentum_2019}
\bibinfo{author}{\bibfnamefont{H.}~\bibnamefont{Pfau}},
  \bibinfo{author}{\bibfnamefont{S.}~\bibnamefont{Chen}},
  \bibinfo{author}{\bibfnamefont{M.}~\bibnamefont{Yi}},
  \bibinfo{author}{\bibfnamefont{M.}~\bibnamefont{Hashimoto}},
  \bibinfo{author}{\bibfnamefont{C.}~\bibnamefont{Rotundu}},
  \bibinfo{author}{\bibfnamefont{J.}~\bibnamefont{Palmstrom}},
  \bibinfo{author}{\bibfnamefont{T.}~\bibnamefont{Chen}},
  \bibinfo{author}{\bibfnamefont{P.-C.} \bibnamefont{Dai}},
  \bibinfo{author}{\bibfnamefont{J.}~\bibnamefont{Straquadine}},
  \bibinfo{author}{\bibfnamefont{A.}~\bibnamefont{Hristov}},
  \bibnamefont{et~al.}, \bibinfo{journal}{Physical Review Letters}
  \textbf{\bibinfo{volume}{123}}, \bibinfo{pages}{066402}
  (\bibinfo{year}{2019}).

\bibitem[{\citenamefont{Sanchez et~al.}(2020)\citenamefont{Sanchez, Malinowski,
  Mutch, Liu, Kim, Ryan, and Chu}}]{sanchez_spontaneous_2020}
\bibinfo{author}{\bibfnamefont{J.~J.} \bibnamefont{Sanchez}},
  \bibinfo{author}{\bibfnamefont{P.}~\bibnamefont{Malinowski}},
  \bibinfo{author}{\bibfnamefont{J.}~\bibnamefont{Mutch}},
  \bibinfo{author}{\bibfnamefont{J.}~\bibnamefont{Liu}},
  \bibinfo{author}{\bibfnamefont{J.-W.} \bibnamefont{Kim}},
  \bibinfo{author}{\bibfnamefont{P.~J.} \bibnamefont{Ryan}}, \bibnamefont{and}
  \bibinfo{author}{\bibfnamefont{J.-H.} \bibnamefont{Chu}},
  \bibinfo{journal}{arXiv:2006.09444 [cond-mat]}  (\bibinfo{year}{2020}).

\bibitem[{\citenamefont{Hristov et~al.}(2019)\citenamefont{Hristov, Ikeda,
  Palmstrom, Walmsley, and Fisher}}]{hristov_elastoresistive_2019}
\bibinfo{author}{\bibfnamefont{A.~T.} \bibnamefont{Hristov}},
  \bibinfo{author}{\bibfnamefont{M.~S.} \bibnamefont{Ikeda}},
  \bibinfo{author}{\bibfnamefont{J.~C.} \bibnamefont{Palmstrom}},
  \bibinfo{author}{\bibfnamefont{P.}~\bibnamefont{Walmsley}}, \bibnamefont{and}
  \bibinfo{author}{\bibfnamefont{I.~R.} \bibnamefont{Fisher}},
  \bibinfo{journal}{Physical Review B} \textbf{\bibinfo{volume}{99}},
  \bibinfo{pages}{100101(R)} (\bibinfo{year}{2019}).

\bibitem[{\citenamefont{Ikeda et~al.}(2020)\citenamefont{Ikeda, Worasaran,
  Rosenberg, Palmstrom, Kivelson, and Fisher}}]{ikeda_elastocaloric_2020}
\bibinfo{author}{\bibfnamefont{M.~S.} \bibnamefont{Ikeda}},
  \bibinfo{author}{\bibfnamefont{T.}~\bibnamefont{Worasaran}},
  \bibinfo{author}{\bibfnamefont{E.~W.} \bibnamefont{Rosenberg}},
  \bibinfo{author}{\bibfnamefont{J.~C.} \bibnamefont{Palmstrom}},
  \bibinfo{author}{\bibfnamefont{S.~A.} \bibnamefont{Kivelson}},
  \bibnamefont{and} \bibinfo{author}{\bibfnamefont{I.~R.}
  \bibnamefont{Fisher}}, \bibinfo{journal}{arXiv:2101.00080 [cond-mat]}
  (\bibinfo{year}{2020}).

\bibitem[{\citenamefont{Caglieris et~al.}(2021)\citenamefont{Caglieris, Wuttke,
  Hong, Sykora, Kappenberger, Aswartham, Wurmehl, Büchner, and
  Hess}}]{caglieris_strain_2021}
\bibinfo{author}{\bibfnamefont{F.}~\bibnamefont{Caglieris}},
  \bibinfo{author}{\bibfnamefont{C.}~\bibnamefont{Wuttke}},
  \bibinfo{author}{\bibfnamefont{X.~C.} \bibnamefont{Hong}},
  \bibinfo{author}{\bibfnamefont{S.}~\bibnamefont{Sykora}},
  \bibinfo{author}{\bibfnamefont{R.}~\bibnamefont{Kappenberger}},
  \bibinfo{author}{\bibfnamefont{S.}~\bibnamefont{Aswartham}},
  \bibinfo{author}{\bibfnamefont{S.}~\bibnamefont{Wurmehl}},
  \bibinfo{author}{\bibfnamefont{B.}~\bibnamefont{Büchner}}, \bibnamefont{and}
  \bibinfo{author}{\bibfnamefont{C.}~\bibnamefont{Hess}}, \bibinfo{journal}{npj
  Quantum Materials} \textbf{\bibinfo{volume}{6}}, \bibinfo{pages}{27}
  (\bibinfo{year}{2021}).

\bibitem[{\citenamefont{Gallais et~al.}(2013)\citenamefont{Gallais, Fernandes,
  Paul, Chauvière, Yang, Méasson, Cazayous, Sacuto, Colson, and
  Forget}}]{gallais_observation_2013}
\bibinfo{author}{\bibfnamefont{Y.}~\bibnamefont{Gallais}},
  \bibinfo{author}{\bibfnamefont{R.~M.} \bibnamefont{Fernandes}},
  \bibinfo{author}{\bibfnamefont{I.}~\bibnamefont{Paul}},
  \bibinfo{author}{\bibfnamefont{L.}~\bibnamefont{Chauvière}},
  \bibinfo{author}{\bibfnamefont{Y.-X.} \bibnamefont{Yang}},
  \bibinfo{author}{\bibfnamefont{M.-A.} \bibnamefont{Méasson}},
  \bibinfo{author}{\bibfnamefont{M.}~\bibnamefont{Cazayous}},
  \bibinfo{author}{\bibfnamefont{A.}~\bibnamefont{Sacuto}},
  \bibinfo{author}{\bibfnamefont{D.}~\bibnamefont{Colson}}, \bibnamefont{and}
  \bibinfo{author}{\bibfnamefont{A.}~\bibnamefont{Forget}},
  \bibinfo{journal}{Physical Review Letters} \textbf{\bibinfo{volume}{111}},
  \bibinfo{pages}{267001} (\bibinfo{year}{2013}).

\bibitem[{\citenamefont{Gallais and Paul}(2016)}]{gallais_charge_2016}
\bibinfo{author}{\bibfnamefont{Y.}~\bibnamefont{Gallais}} \bibnamefont{and}
  \bibinfo{author}{\bibfnamefont{I.}~\bibnamefont{Paul}},
  \bibinfo{journal}{Comptes Rendus Physique} \textbf{\bibinfo{volume}{17}},
  \bibinfo{pages}{113} (\bibinfo{year}{2016}).

\bibitem[{\citenamefont{Thorsmølle et~al.}(2016)\citenamefont{Thorsmølle,
  Khodas, Yin, Zhang, Carr, Dai, and Blumberg}}]{thorsmolle_critical_2016}
\bibinfo{author}{\bibfnamefont{V.~K.} \bibnamefont{Thorsmølle}},
  \bibinfo{author}{\bibfnamefont{M.}~\bibnamefont{Khodas}},
  \bibinfo{author}{\bibfnamefont{Z.~P.} \bibnamefont{Yin}},
  \bibinfo{author}{\bibfnamefont{C.}~\bibnamefont{Zhang}},
  \bibinfo{author}{\bibfnamefont{S.~V.} \bibnamefont{Carr}},
  \bibinfo{author}{\bibfnamefont{P.}~\bibnamefont{Dai}}, \bibnamefont{and}
  \bibinfo{author}{\bibfnamefont{G.}~\bibnamefont{Blumberg}},
  \bibinfo{journal}{Physical Review B} \textbf{\bibinfo{volume}{93}},
  \bibinfo{pages}{054515} (\bibinfo{year}{2016}).

\bibitem[{\citenamefont{Kretzschmar et~al.}(2016)\citenamefont{Kretzschmar,
  Böhm, Karahasanović, Muschler, Baum, Jost, Schmalian, Caprara, Grilli,
  Di~Castro et~al.}}]{kretzschmar_critical_2016}
\bibinfo{author}{\bibfnamefont{F.}~\bibnamefont{Kretzschmar}},
  \bibinfo{author}{\bibfnamefont{T.}~\bibnamefont{Böhm}},
  \bibinfo{author}{\bibfnamefont{U.}~\bibnamefont{Karahasanović}},
  \bibinfo{author}{\bibfnamefont{B.}~\bibnamefont{Muschler}},
  \bibinfo{author}{\bibfnamefont{A.}~\bibnamefont{Baum}},
  \bibinfo{author}{\bibfnamefont{D.}~\bibnamefont{Jost}},
  \bibinfo{author}{\bibfnamefont{J.}~\bibnamefont{Schmalian}},
  \bibinfo{author}{\bibfnamefont{S.}~\bibnamefont{Caprara}},
  \bibinfo{author}{\bibfnamefont{M.}~\bibnamefont{Grilli}},
  \bibinfo{author}{\bibfnamefont{C.}~\bibnamefont{Di~Castro}},
  \bibnamefont{et~al.}, \bibinfo{journal}{Nature Physics}
  \textbf{\bibinfo{volume}{12}}, \bibinfo{pages}{560} (\bibinfo{year}{2016}).

\bibitem[{\citenamefont{Massat et~al.}(2016)\citenamefont{Massat, Farina, Paul,
  Karlsson, Strobel, Toulemonde, Méasson, Cazayous, Sacuto, Kasahara
  et~al.}}]{massat_charge-induced_2016}
\bibinfo{author}{\bibfnamefont{P.}~\bibnamefont{Massat}},
  \bibinfo{author}{\bibfnamefont{D.}~\bibnamefont{Farina}},
  \bibinfo{author}{\bibfnamefont{I.}~\bibnamefont{Paul}},
  \bibinfo{author}{\bibfnamefont{S.}~\bibnamefont{Karlsson}},
  \bibinfo{author}{\bibfnamefont{P.}~\bibnamefont{Strobel}},
  \bibinfo{author}{\bibfnamefont{P.}~\bibnamefont{Toulemonde}},
  \bibinfo{author}{\bibfnamefont{M.-A.} \bibnamefont{Méasson}},
  \bibinfo{author}{\bibfnamefont{M.}~\bibnamefont{Cazayous}},
  \bibinfo{author}{\bibfnamefont{A.}~\bibnamefont{Sacuto}},
  \bibinfo{author}{\bibfnamefont{S.}~\bibnamefont{Kasahara}},
  \bibnamefont{et~al.}, \bibinfo{journal}{Proceedings of the National Academy
  of Sciences} \textbf{\bibinfo{volume}{113}}, \bibinfo{pages}{9177}
  (\bibinfo{year}{2016}).

\bibitem[{\citenamefont{Ren et~al.}(2015)\citenamefont{Ren, Duan, Hu, Li,
  Zhang, Luo, Dai, and Li}}]{ren_nematic_2015}
\bibinfo{author}{\bibfnamefont{X.}~\bibnamefont{Ren}},
  \bibinfo{author}{\bibfnamefont{L.}~\bibnamefont{Duan}},
  \bibinfo{author}{\bibfnamefont{Y.}~\bibnamefont{Hu}},
  \bibinfo{author}{\bibfnamefont{J.}~\bibnamefont{Li}},
  \bibinfo{author}{\bibfnamefont{R.}~\bibnamefont{Zhang}},
  \bibinfo{author}{\bibfnamefont{H.}~\bibnamefont{Luo}},
  \bibinfo{author}{\bibfnamefont{P.}~\bibnamefont{Dai}}, \bibnamefont{and}
  \bibinfo{author}{\bibfnamefont{Y.}~\bibnamefont{Li}},
  \bibinfo{journal}{Physical Review Letters} \textbf{\bibinfo{volume}{115}},
  \bibinfo{pages}{197002} (\bibinfo{year}{2015}).

\bibitem[{\citenamefont{Baum et~al.}(2018)\citenamefont{Baum, Li, Tomić,
  Lazarević, Jost, Löffler, Muschler, Böhm, Chu, Fisher
  et~al.}}]{baum_interplay_2018}
\bibinfo{author}{\bibfnamefont{A.}~\bibnamefont{Baum}},
  \bibinfo{author}{\bibfnamefont{Y.}~\bibnamefont{Li}},
  \bibinfo{author}{\bibfnamefont{M.}~\bibnamefont{Tomić}},
  \bibinfo{author}{\bibfnamefont{N.}~\bibnamefont{Lazarević}},
  \bibinfo{author}{\bibfnamefont{D.}~\bibnamefont{Jost}},
  \bibinfo{author}{\bibfnamefont{F.}~\bibnamefont{Löffler}},
  \bibinfo{author}{\bibfnamefont{B.}~\bibnamefont{Muschler}},
  \bibinfo{author}{\bibfnamefont{T.}~\bibnamefont{Böhm}},
  \bibinfo{author}{\bibfnamefont{J.-H.} \bibnamefont{Chu}},
  \bibinfo{author}{\bibfnamefont{I.~R.} \bibnamefont{Fisher}},
  \bibnamefont{et~al.}, \bibinfo{journal}{Physical Review B}
  \textbf{\bibinfo{volume}{98}}, \bibinfo{pages}{075113}
  (\bibinfo{year}{2018}).

\bibitem[{\citenamefont{Fleury and
  Worlock}(1968)}]{fleury_electric-field-induced_1968}
\bibinfo{author}{\bibfnamefont{P.~A.} \bibnamefont{Fleury}} \bibnamefont{and}
  \bibinfo{author}{\bibfnamefont{J.~M.} \bibnamefont{Worlock}},
  \bibinfo{journal}{Physical Review} \textbf{\bibinfo{volume}{174}},
  \bibinfo{pages}{613} (\bibinfo{year}{1968}).

\bibitem[{\citenamefont{Anastassakis and
  Burstein}(1971)}]{anastassakis_morphic_1971}
\bibinfo{author}{\bibfnamefont{E.}~\bibnamefont{Anastassakis}}
  \bibnamefont{and} \bibinfo{author}{\bibfnamefont{E.}~\bibnamefont{Burstein}},
  \bibinfo{journal}{Journal of Physics and Chemistry of Solids}
  \textbf{\bibinfo{volume}{32}}, \bibinfo{pages}{563} (\bibinfo{year}{1971}).

\bibitem[{\citenamefont{Ganesan et~al.}(1970)\citenamefont{Ganesan, Maradudin,
  and Oitmaa}}]{ganesan_lattice_1970}
\bibinfo{author}{\bibfnamefont{S.}~\bibnamefont{Ganesan}},
  \bibinfo{author}{\bibfnamefont{A.~A.} \bibnamefont{Maradudin}},
  \bibnamefont{and} \bibinfo{author}{\bibfnamefont{J.}~\bibnamefont{Oitmaa}},
  \bibinfo{journal}{Annals of Physics} \textbf{\bibinfo{volume}{56}},
  \bibinfo{pages}{556} (\bibinfo{year}{1970}).

\bibitem[{\citenamefont{Melo and Cerdeira}(1982)}]{melo_changes_1982}
\bibinfo{author}{\bibfnamefont{F.~E.~A.} \bibnamefont{Melo}} \bibnamefont{and}
  \bibinfo{author}{\bibfnamefont{F.}~\bibnamefont{Cerdeira}},
  \bibinfo{journal}{Physical Review B} \textbf{\bibinfo{volume}{26}},
  \bibinfo{pages}{720} (\bibinfo{year}{1982}).

\bibitem[{\citenamefont{Merle et~al.}(1980)\citenamefont{Merle, Pascual,
  Camassel, and Mathieu}}]{merle_uniaxial-stress_1980}
\bibinfo{author}{\bibfnamefont{P.}~\bibnamefont{Merle}},
  \bibinfo{author}{\bibfnamefont{J.}~\bibnamefont{Pascual}},
  \bibinfo{author}{\bibfnamefont{J.}~\bibnamefont{Camassel}}, \bibnamefont{and}
  \bibinfo{author}{\bibfnamefont{H.}~\bibnamefont{Mathieu}},
  \bibinfo{journal}{Physical Review B} \textbf{\bibinfo{volume}{21}},
  \bibinfo{pages}{1617} (\bibinfo{year}{1980}).

\bibitem[{\citenamefont{Anastassakis et~al.}(1993)\citenamefont{Anastassakis,
  Pinczuk, Burstein, Pollak, and Cardona}}]{anastassakis_effect_1993}
\bibinfo{author}{\bibfnamefont{E.}~\bibnamefont{Anastassakis}},
  \bibinfo{author}{\bibfnamefont{A.}~\bibnamefont{Pinczuk}},
  \bibinfo{author}{\bibfnamefont{E.}~\bibnamefont{Burstein}},
  \bibinfo{author}{\bibfnamefont{F.~H.} \bibnamefont{Pollak}},
  \bibnamefont{and} \bibinfo{author}{\bibfnamefont{M.}~\bibnamefont{Cardona}},
  \bibinfo{journal}{Solid State Communications} \textbf{\bibinfo{volume}{88}},
  \bibinfo{pages}{1053} (\bibinfo{year}{1993}).

\bibitem[{\citenamefont{Ikeda et~al.}(2018)\citenamefont{Ikeda, Worasaran,
  Palmstrom, Straquadine, Walmsley, and Fisher}}]{ikeda_symmetric_2018}
\bibinfo{author}{\bibfnamefont{M.~S.} \bibnamefont{Ikeda}},
  \bibinfo{author}{\bibfnamefont{T.}~\bibnamefont{Worasaran}},
  \bibinfo{author}{\bibfnamefont{J.~C.} \bibnamefont{Palmstrom}},
  \bibinfo{author}{\bibfnamefont{J.~A.~W.} \bibnamefont{Straquadine}},
  \bibinfo{author}{\bibfnamefont{P.}~\bibnamefont{Walmsley}}, \bibnamefont{and}
  \bibinfo{author}{\bibfnamefont{I.~R.} \bibnamefont{Fisher}},
  \bibinfo{journal}{Physical Review B} \textbf{\bibinfo{volume}{98}},
  \bibinfo{pages}{245133} (\bibinfo{year}{2018}).

\bibitem[{\citenamefont{Chauvière et~al.}(2011)\citenamefont{Chauvière,
  Gallais, Cazayous, Méasson, Sacuto, Colson, and
  Forget}}]{chauviere_raman_2011}
\bibinfo{author}{\bibfnamefont{L.}~\bibnamefont{Chauvière}},
  \bibinfo{author}{\bibfnamefont{Y.}~\bibnamefont{Gallais}},
  \bibinfo{author}{\bibfnamefont{M.}~\bibnamefont{Cazayous}},
  \bibinfo{author}{\bibfnamefont{M.~A.} \bibnamefont{Méasson}},
  \bibinfo{author}{\bibfnamefont{A.}~\bibnamefont{Sacuto}},
  \bibinfo{author}{\bibfnamefont{D.}~\bibnamefont{Colson}}, \bibnamefont{and}
  \bibinfo{author}{\bibfnamefont{A.}~\bibnamefont{Forget}},
  \bibinfo{journal}{Physical Review B} \textbf{\bibinfo{volume}{84}},
  \bibinfo{pages}{104508} (\bibinfo{year}{2011}).

\bibitem[{\citenamefont{Kuroki et~al.}(2009)\citenamefont{Kuroki, Usui, Onari,
  Arita, and Aoki}}]{kuroki_pnictogen_2009}
\bibinfo{author}{\bibfnamefont{K.}~\bibnamefont{Kuroki}},
  \bibinfo{author}{\bibfnamefont{H.}~\bibnamefont{Usui}},
  \bibinfo{author}{\bibfnamefont{S.}~\bibnamefont{Onari}},
  \bibinfo{author}{\bibfnamefont{R.}~\bibnamefont{Arita}}, \bibnamefont{and}
  \bibinfo{author}{\bibfnamefont{H.}~\bibnamefont{Aoki}},
  \bibinfo{journal}{Physical Review B} \textbf{\bibinfo{volume}{79}},
  \bibinfo{pages}{224511} (\bibinfo{year}{2009}).

\bibitem[{\citenamefont{Yildirim}(2009)}]{yildirim_strong_2009}
\bibinfo{author}{\bibfnamefont{T.}~\bibnamefont{Yildirim}},
  \bibinfo{journal}{Physical Review Letters} \textbf{\bibinfo{volume}{102}},
  \bibinfo{pages}{037003} (\bibinfo{year}{2009}).

\bibitem[{\citenamefont{Yndurain and Soler}(2009)}]{yndurain_anomalous_2009}
\bibinfo{author}{\bibfnamefont{F.}~\bibnamefont{Yndurain}} \bibnamefont{and}
  \bibinfo{author}{\bibfnamefont{J.~M.} \bibnamefont{Soler}},
  \bibinfo{journal}{Physical Review B} \textbf{\bibinfo{volume}{79}},
  \bibinfo{pages}{134506} (\bibinfo{year}{2009}).

\bibitem[{\citenamefont{Lee et~al.}(2008)\citenamefont{Lee, Iyo, Eisaki, Kito,
  Teresa Fernandez-Diaz, Ito, Kihou, Matsuhata, Braden, and
  Yamada}}]{lee_effect_2008}
\bibinfo{author}{\bibfnamefont{C.-H.} \bibnamefont{Lee}},
  \bibinfo{author}{\bibfnamefont{A.}~\bibnamefont{Iyo}},
  \bibinfo{author}{\bibfnamefont{H.}~\bibnamefont{Eisaki}},
  \bibinfo{author}{\bibfnamefont{H.}~\bibnamefont{Kito}},
  \bibinfo{author}{\bibfnamefont{M.}~\bibnamefont{Teresa Fernandez-Diaz}},
  \bibinfo{author}{\bibfnamefont{T.}~\bibnamefont{Ito}},
  \bibinfo{author}{\bibfnamefont{K.}~\bibnamefont{Kihou}},
  \bibinfo{author}{\bibfnamefont{H.}~\bibnamefont{Matsuhata}},
  \bibinfo{author}{\bibfnamefont{M.}~\bibnamefont{Braden}}, \bibnamefont{and}
  \bibinfo{author}{\bibfnamefont{K.}~\bibnamefont{Yamada}},
  \bibinfo{journal}{Journal of the Physical Society of Japan}
  \textbf{\bibinfo{volume}{77}}, \bibinfo{pages}{083704}
  (\bibinfo{year}{2008}).

\bibitem[{\citenamefont{Wu et~al.}(2020{\natexlab{a}})\citenamefont{Wu, Zhang,
  Thorsmølle, Chen, Tan, Dai, Shi, Jin, Shibauchi, Kasahara
  et~al.}}]{wu_-plane_2020}
\bibinfo{author}{\bibfnamefont{S.-F.} \bibnamefont{Wu}},
  \bibinfo{author}{\bibfnamefont{W.-L.} \bibnamefont{Zhang}},
  \bibinfo{author}{\bibfnamefont{V.~K.} \bibnamefont{Thorsmølle}},
  \bibinfo{author}{\bibfnamefont{G.~F.} \bibnamefont{Chen}},
  \bibinfo{author}{\bibfnamefont{G.~T.} \bibnamefont{Tan}},
  \bibinfo{author}{\bibfnamefont{P.~C.} \bibnamefont{Dai}},
  \bibinfo{author}{\bibfnamefont{Y.~G.} \bibnamefont{Shi}},
  \bibinfo{author}{\bibfnamefont{C.~Q.} \bibnamefont{Jin}},
  \bibinfo{author}{\bibfnamefont{T.}~\bibnamefont{Shibauchi}},
  \bibinfo{author}{\bibfnamefont{S.}~\bibnamefont{Kasahara}},
  \bibnamefont{et~al.}, \bibinfo{journal}{Physical Review Research}
  \textbf{\bibinfo{volume}{2}}, \bibinfo{pages}{033140}
  (\bibinfo{year}{2020}{\natexlab{a}}).

\bibitem[{\citenamefont{García-Martínez
  et~al.}(2013)\citenamefont{García-Martínez, Valenzuela, Ciuchi, Cappelluti,
  Calderón, and Bascones}}]{garcia-martinez_coupling_2013}
\bibinfo{author}{\bibfnamefont{N.~A.} \bibnamefont{García-Martínez}},
  \bibinfo{author}{\bibfnamefont{B.}~\bibnamefont{Valenzuela}},
  \bibinfo{author}{\bibfnamefont{S.}~\bibnamefont{Ciuchi}},
  \bibinfo{author}{\bibfnamefont{E.}~\bibnamefont{Cappelluti}},
  \bibinfo{author}{\bibfnamefont{M.~J.} \bibnamefont{Calderón}},
  \bibnamefont{and} \bibinfo{author}{\bibfnamefont{E.}~\bibnamefont{Bascones}},
  \bibinfo{journal}{Physical Review B} \textbf{\bibinfo{volume}{88}},
  \bibinfo{pages}{165106} (\bibinfo{year}{2013}).

\bibitem[{\citenamefont{Rotter et~al.}(2008)\citenamefont{Rotter, Tegel,
  Johrendt, Schellenberg, Hermes, and
  Pöttgen}}]{rotter_spin-density-wave_2008}
\bibinfo{author}{\bibfnamefont{M.}~\bibnamefont{Rotter}},
  \bibinfo{author}{\bibfnamefont{M.}~\bibnamefont{Tegel}},
  \bibinfo{author}{\bibfnamefont{D.}~\bibnamefont{Johrendt}},
  \bibinfo{author}{\bibfnamefont{I.}~\bibnamefont{Schellenberg}},
  \bibinfo{author}{\bibfnamefont{W.}~\bibnamefont{Hermes}}, \bibnamefont{and}
  \bibinfo{author}{\bibfnamefont{R.}~\bibnamefont{Pöttgen}},
  \bibinfo{journal}{Physical Review B} \textbf{\bibinfo{volume}{78}},
  \bibinfo{pages}{020503} (\bibinfo{year}{2008}).

\bibitem[{\citenamefont{Yoshizawa et~al.}(2012)\citenamefont{Yoshizawa, Kimura,
  Chiba, Ismayil, Nakanishi, Kihou, Lee, Iyo, Eisaki, Nakajima
  et~al.}}]{yoshizawa_structural_2012}
\bibinfo{author}{\bibfnamefont{M.}~\bibnamefont{Yoshizawa}},
  \bibinfo{author}{\bibfnamefont{D.}~\bibnamefont{Kimura}},
  \bibinfo{author}{\bibfnamefont{T.}~\bibnamefont{Chiba}},
  \bibinfo{author}{\bibfnamefont{A.}~\bibnamefont{Ismayil}},
  \bibinfo{author}{\bibfnamefont{Y.}~\bibnamefont{Nakanishi}},
  \bibinfo{author}{\bibfnamefont{K.}~\bibnamefont{Kihou}},
  \bibinfo{author}{\bibfnamefont{C.-H.} \bibnamefont{Lee}},
  \bibinfo{author}{\bibfnamefont{A.}~\bibnamefont{Iyo}},
  \bibinfo{author}{\bibfnamefont{H.}~\bibnamefont{Eisaki}},
  \bibinfo{author}{\bibfnamefont{M.}~\bibnamefont{Nakajima}},
  \bibnamefont{et~al.}, \bibinfo{journal}{Journal of the Physical Society of
  Japan} \textbf{\bibinfo{volume}{81}}, \bibinfo{pages}{024604}
  (\bibinfo{year}{2012}).

\bibitem[{\citenamefont{Yoshizawa and Simayi}(2012)}]{yoshizawa_anomalous_2012}
\bibinfo{author}{\bibfnamefont{M.}~\bibnamefont{Yoshizawa}} \bibnamefont{and}
  \bibinfo{author}{\bibfnamefont{S.}~\bibnamefont{Simayi}},
  \bibinfo{journal}{Modern Physics Letters B} \textbf{\bibinfo{volume}{26}},
  \bibinfo{pages}{1230011} (\bibinfo{year}{2012}).

\bibitem[{\citenamefont{Böhmer and Meingast}(2016)}]{bohmer_electronic_2016}
\bibinfo{author}{\bibfnamefont{A.~E.} \bibnamefont{Böhmer}} \bibnamefont{and}
  \bibinfo{author}{\bibfnamefont{C.}~\bibnamefont{Meingast}},
  \bibinfo{journal}{Comptes Rendus Physique} \textbf{\bibinfo{volume}{17}},
  \bibinfo{pages}{90} (\bibinfo{year}{2016}).

\bibitem[{\citenamefont{Dhital et~al.}(2012)\citenamefont{Dhital, Yamani, Tian,
  Zeretsky, Sefat, Wang, Birgeneau, and Wilson}}]{dhital_effect_2012}
\bibinfo{author}{\bibfnamefont{C.}~\bibnamefont{Dhital}},
  \bibinfo{author}{\bibfnamefont{Z.}~\bibnamefont{Yamani}},
  \bibinfo{author}{\bibfnamefont{W.}~\bibnamefont{Tian}},
  \bibinfo{author}{\bibfnamefont{J.}~\bibnamefont{Zeretsky}},
  \bibinfo{author}{\bibfnamefont{A.~S.} \bibnamefont{Sefat}},
  \bibinfo{author}{\bibfnamefont{Z.}~\bibnamefont{Wang}},
  \bibinfo{author}{\bibfnamefont{R.~J.} \bibnamefont{Birgeneau}},
  \bibnamefont{and} \bibinfo{author}{\bibfnamefont{S.~D.}
  \bibnamefont{Wilson}}, \bibinfo{journal}{Physical Review Letters}
  \textbf{\bibinfo{volume}{108}}, \bibinfo{pages}{087001}
  (\bibinfo{year}{2012}).

\bibitem[{\citenamefont{Lu et~al.}(2016)\citenamefont{Lu, Tseng, Keller, Zhang,
  Hu, Song, Man, Park, Luo, Li et~al.}}]{lu_impact_2016}
\bibinfo{author}{\bibfnamefont{X.}~\bibnamefont{Lu}},
  \bibinfo{author}{\bibfnamefont{K.-F.} \bibnamefont{Tseng}},
  \bibinfo{author}{\bibfnamefont{T.}~\bibnamefont{Keller}},
  \bibinfo{author}{\bibfnamefont{W.}~\bibnamefont{Zhang}},
  \bibinfo{author}{\bibfnamefont{D.}~\bibnamefont{Hu}},
  \bibinfo{author}{\bibfnamefont{Y.}~\bibnamefont{Song}},
  \bibinfo{author}{\bibfnamefont{H.}~\bibnamefont{Man}},
  \bibinfo{author}{\bibfnamefont{J.~T.} \bibnamefont{Park}},
  \bibinfo{author}{\bibfnamefont{H.}~\bibnamefont{Luo}},
  \bibinfo{author}{\bibfnamefont{S.}~\bibnamefont{Li}}, \bibnamefont{et~al.},
  \bibinfo{journal}{Physical Review B} \textbf{\bibinfo{volume}{93}},
  \bibinfo{pages}{134519} (\bibinfo{year}{2016}).

\bibitem[{\citenamefont{Tam et~al.}(2017)\citenamefont{Tam, Song, Man, Cheung,
  Yin, Lu, Wang, Frandsen, Liu, Gong et~al.}}]{tam_uniaxial_2017}
\bibinfo{author}{\bibfnamefont{D.~W.} \bibnamefont{Tam}},
  \bibinfo{author}{\bibfnamefont{Y.}~\bibnamefont{Song}},
  \bibinfo{author}{\bibfnamefont{H.}~\bibnamefont{Man}},
  \bibinfo{author}{\bibfnamefont{S.~C.} \bibnamefont{Cheung}},
  \bibinfo{author}{\bibfnamefont{Z.}~\bibnamefont{Yin}},
  \bibinfo{author}{\bibfnamefont{X.}~\bibnamefont{Lu}},
  \bibinfo{author}{\bibfnamefont{W.}~\bibnamefont{Wang}},
  \bibinfo{author}{\bibfnamefont{B.~A.} \bibnamefont{Frandsen}},
  \bibinfo{author}{\bibfnamefont{L.}~\bibnamefont{Liu}},
  \bibinfo{author}{\bibfnamefont{Z.}~\bibnamefont{Gong}}, \bibnamefont{et~al.},
  \bibinfo{journal}{Physical Review B} \textbf{\bibinfo{volume}{95}},
  \bibinfo{pages}{060505} (\bibinfo{year}{2017}).

\bibitem[{\citenamefont{Chauvière et~al.}(2009)\citenamefont{Chauvière,
  Gallais, Cazayous, Sacuto, Méasson, Colson, and
  Forget}}]{chauviere_doping_2009}
\bibinfo{author}{\bibfnamefont{L.}~\bibnamefont{Chauvière}},
  \bibinfo{author}{\bibfnamefont{Y.}~\bibnamefont{Gallais}},
  \bibinfo{author}{\bibfnamefont{M.}~\bibnamefont{Cazayous}},
  \bibinfo{author}{\bibfnamefont{A.}~\bibnamefont{Sacuto}},
  \bibinfo{author}{\bibfnamefont{M.~A.} \bibnamefont{Méasson}},
  \bibinfo{author}{\bibfnamefont{D.}~\bibnamefont{Colson}}, \bibnamefont{and}
  \bibinfo{author}{\bibfnamefont{A.}~\bibnamefont{Forget}},
  \bibinfo{journal}{Physical Review B} \textbf{\bibinfo{volume}{80}},
  \bibinfo{pages}{094504} (\bibinfo{year}{2009}).

\bibitem[{\citenamefont{Wu et~al.}(2020{\natexlab{b}})\citenamefont{Wu, Zhang,
  Li, Cao, Kung, Sefat, Ding, Richard, and Blumberg}}]{wu_coupling_2020}
\bibinfo{author}{\bibfnamefont{S.-F.} \bibnamefont{Wu}},
  \bibinfo{author}{\bibfnamefont{W.-L.} \bibnamefont{Zhang}},
  \bibinfo{author}{\bibfnamefont{L.}~\bibnamefont{Li}},
  \bibinfo{author}{\bibfnamefont{H.-B.} \bibnamefont{Cao}},
  \bibinfo{author}{\bibfnamefont{H.-H.} \bibnamefont{Kung}},
  \bibinfo{author}{\bibfnamefont{A.~S.} \bibnamefont{Sefat}},
  \bibinfo{author}{\bibfnamefont{H.}~\bibnamefont{Ding}},
  \bibinfo{author}{\bibfnamefont{P.}~\bibnamefont{Richard}}, \bibnamefont{and}
  \bibinfo{author}{\bibfnamefont{G.}~\bibnamefont{Blumberg}},
  \bibinfo{journal}{Physical Review B} \textbf{\bibinfo{volume}{102}},
  \bibinfo{pages}{014501} (\bibinfo{year}{2020}{\natexlab{b}}).

\bibitem[{\citenamefont{Klein}(1983)}]{klein_electronic_1983}
\bibinfo{author}{\bibfnamefont{M.~V.} \bibnamefont{Klein}},
  \emph{\bibinfo{title}{Electronic {Raman} scattering}},
  vol.~\bibinfo{volume}{8} of \emph{\bibinfo{series}{Topics in {Applied}
  {Physics}}} (\bibinfo{publisher}{Springer Berlin-Heidelberg},
  \bibinfo{year}{1983}).

\bibitem[{\citenamefont{Rahlenbeck et~al.}(2009)\citenamefont{Rahlenbeck, Sun,
  Sun, Lin, Keimer, and Ulrich}}]{rahlenbeck_phonon_2009}
\bibinfo{author}{\bibfnamefont{M.}~\bibnamefont{Rahlenbeck}},
  \bibinfo{author}{\bibfnamefont{G.}~\bibnamefont{Sun}},
  \bibinfo{author}{\bibfnamefont{D.}~\bibnamefont{Sun}},
  \bibinfo{author}{\bibfnamefont{C.}~\bibnamefont{Lin}},
  \bibinfo{author}{\bibfnamefont{B.}~\bibnamefont{Keimer}}, \bibnamefont{and}
  \bibinfo{author}{\bibfnamefont{C.}~\bibnamefont{Ulrich}},
  \bibinfo{journal}{Physical Review B} \textbf{\bibinfo{volume}{80}},
  \bibinfo{pages}{064509} (\bibinfo{year}{2009}).

\bibitem[{\citenamefont{Böhmer et~al.}(2014)\citenamefont{Böhmer, Burger,
  Hardy, Wolf, Schweiss, Fromknecht, Reinecker, Schranz, and
  Meingast}}]{bohmer_nematic_2014}
\bibinfo{author}{\bibfnamefont{A.}~\bibnamefont{Böhmer}},
  \bibinfo{author}{\bibfnamefont{P.}~\bibnamefont{Burger}},
  \bibinfo{author}{\bibfnamefont{F.}~\bibnamefont{Hardy}},
  \bibinfo{author}{\bibfnamefont{T.}~\bibnamefont{Wolf}},
  \bibinfo{author}{\bibfnamefont{P.}~\bibnamefont{Schweiss}},
  \bibinfo{author}{\bibfnamefont{R.}~\bibnamefont{Fromknecht}},
  \bibinfo{author}{\bibfnamefont{M.}~\bibnamefont{Reinecker}},
  \bibinfo{author}{\bibfnamefont{W.}~\bibnamefont{Schranz}}, \bibnamefont{and}
  \bibinfo{author}{\bibfnamefont{C.}~\bibnamefont{Meingast}},
  \bibinfo{journal}{Physical Review Letters} \textbf{\bibinfo{volume}{112}},
  \bibinfo{pages}{047001} (\bibinfo{year}{2014}).

\bibitem[{\citenamefont{Barber}(2018)}]{barber_thesis_2018}
\bibinfo{author}{\bibfnamefont{M.~E.} \bibnamefont{Barber}}, Ph.D. thesis,
  \bibinfo{school}{University of St Andrews, United Kingdom}
  (\bibinfo{year}{2018}).

\bibitem[{\citenamefont{Fujii et~al.}(2018)\citenamefont{Fujii, Simayi, Sakano,
  Sasaki, Nakamura, Nakanishi, Kihou, Nakajima, Lee, Iyo
  et~al.}}]{fujii_diverse_2018}
\bibinfo{author}{\bibfnamefont{C.}~\bibnamefont{Fujii}},
  \bibinfo{author}{\bibfnamefont{S.}~\bibnamefont{Simayi}},
  \bibinfo{author}{\bibfnamefont{K.}~\bibnamefont{Sakano}},
  \bibinfo{author}{\bibfnamefont{C.}~\bibnamefont{Sasaki}},
  \bibinfo{author}{\bibfnamefont{M.}~\bibnamefont{Nakamura}},
  \bibinfo{author}{\bibfnamefont{Y.}~\bibnamefont{Nakanishi}},
  \bibinfo{author}{\bibfnamefont{K.}~\bibnamefont{Kihou}},
  \bibinfo{author}{\bibfnamefont{M.}~\bibnamefont{Nakajima}},
  \bibinfo{author}{\bibfnamefont{C.-H.} \bibnamefont{Lee}},
  \bibinfo{author}{\bibfnamefont{A.}~\bibnamefont{Iyo}}, \bibnamefont{et~al.},
  \bibinfo{journal}{Journal of the Physical Society of Japan}
  \textbf{\bibinfo{volume}{87}}, \bibinfo{pages}{074710}
  (\bibinfo{year}{2018}).

\bibitem[{\citenamefont{Fernandes et~al.}(2010)\citenamefont{Fernandes,
  VanBebber, Bhattacharya, Chandra, Keppens, Mandrus, McGuire, Sales, Sefat,
  and Schmalian}}]{fernandes_effects_2010}
\bibinfo{author}{\bibfnamefont{R.~M.} \bibnamefont{Fernandes}},
  \bibinfo{author}{\bibfnamefont{L.~H.} \bibnamefont{VanBebber}},
  \bibinfo{author}{\bibfnamefont{S.}~\bibnamefont{Bhattacharya}},
  \bibinfo{author}{\bibfnamefont{P.}~\bibnamefont{Chandra}},
  \bibinfo{author}{\bibfnamefont{V.}~\bibnamefont{Keppens}},
  \bibinfo{author}{\bibfnamefont{D.}~\bibnamefont{Mandrus}},
  \bibinfo{author}{\bibfnamefont{M.~A.} \bibnamefont{McGuire}},
  \bibinfo{author}{\bibfnamefont{B.~C.} \bibnamefont{Sales}},
  \bibinfo{author}{\bibfnamefont{A.~S.} \bibnamefont{Sefat}}, \bibnamefont{and}
  \bibinfo{author}{\bibfnamefont{J.}~\bibnamefont{Schmalian}},
  \bibinfo{journal}{Physical Review Letters} \textbf{\bibinfo{volume}{105}},
  \bibinfo{pages}{157003} (\bibinfo{year}{2010}).

\end{thebibliography}
\pagebreak
\appendix

\section{Fano lineshape fitting}
\label{Fano lineshape fitting}
Close to $T_S$ the Raman As phonon displays a significant asymmetry which is linked to its coupling to electronic degrees of freedom. The coupling makes the extraction of the phonon intensity non-trivial, requiring a fit to disentangle the phononic and electronic contribution. The coupling between a discrete mode, here the As phonon, and a broad continuum, here the electronic excitations, leads to a characteristic asymmetrical lineshape which can be described by a coupled mode analysis as depicted in the Fano model. Following Klein's approach \cite{klein_electronic_1983} we consider a bare phonon mode, expressed by a Dirac $\delta$ function with area $\pi t_{ph}$ coupled to a featureless electronic continuum $\rho$ which couples to light with an amplitude $t_e$. The imaginary part of the coupled electron-phonon response $\chi$, the quantity that is measured in a Raman scattering experiment, takes the form:
\begin{equation}
\chi''(\omega)=\frac{t_e^2\pi\rho(\omega_0-\omega-vt_{ph}/t_e)^2}{(\omega_0-\omega)^2+(v^2\pi\rho)^2}
\end{equation}
where $v$ is the electron-phonon coupling parameter and $\omega_0$ is a renormalized effective phonon energy \cite{klein_electronic_1983}.
For the As phonon fitting we used the following parametrization: $A=\pi \rho t_e^2$, $\omega_0$, $t_{ph}$ and $B=\frac{v}{t_e}$. This parametrization has the advantage of involving explicitely the quantity of interest, the bare phonon amplitude $t_{ph}$, in the minimization routine, thus reducing the standard error on its evaluation. The response then takes the form:
\begin{equation}
\chi''(\omega)=A\frac{(\omega_0-\omega-t_{ph}B)^2}{(\omega_0-\omega)^2+(AB^2)^2}
\end{equation}
\par

\begin{figure}[]
\centering
\includegraphics[scale=0.2]{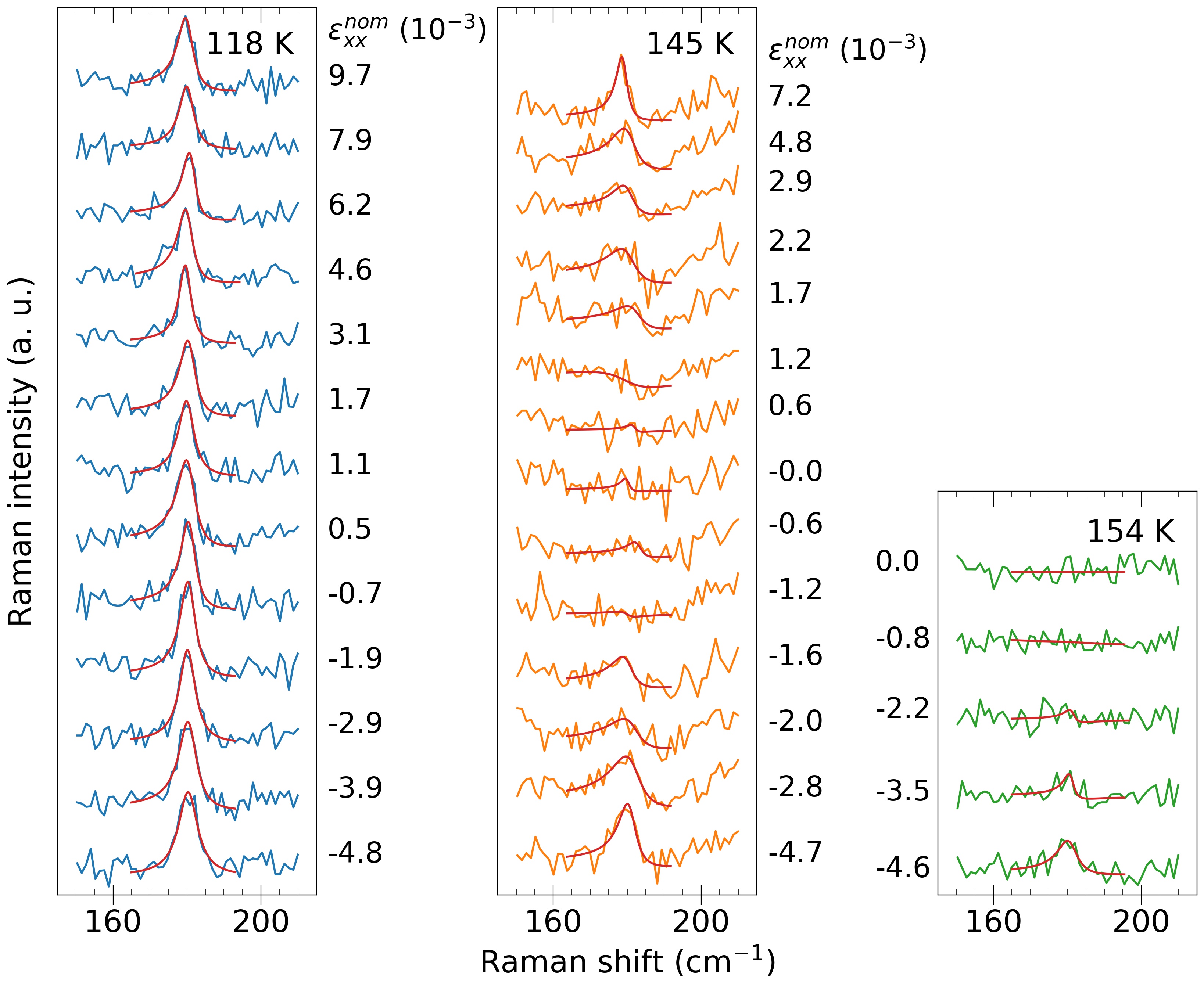}
\caption{Fano lineshape fitting of the As phonon as a function of applied strain and at three different temperatures.}
\label{Strain-dep-fit}
\end{figure}

\begin{figure}[]
\centering
\includegraphics[scale=0.3]{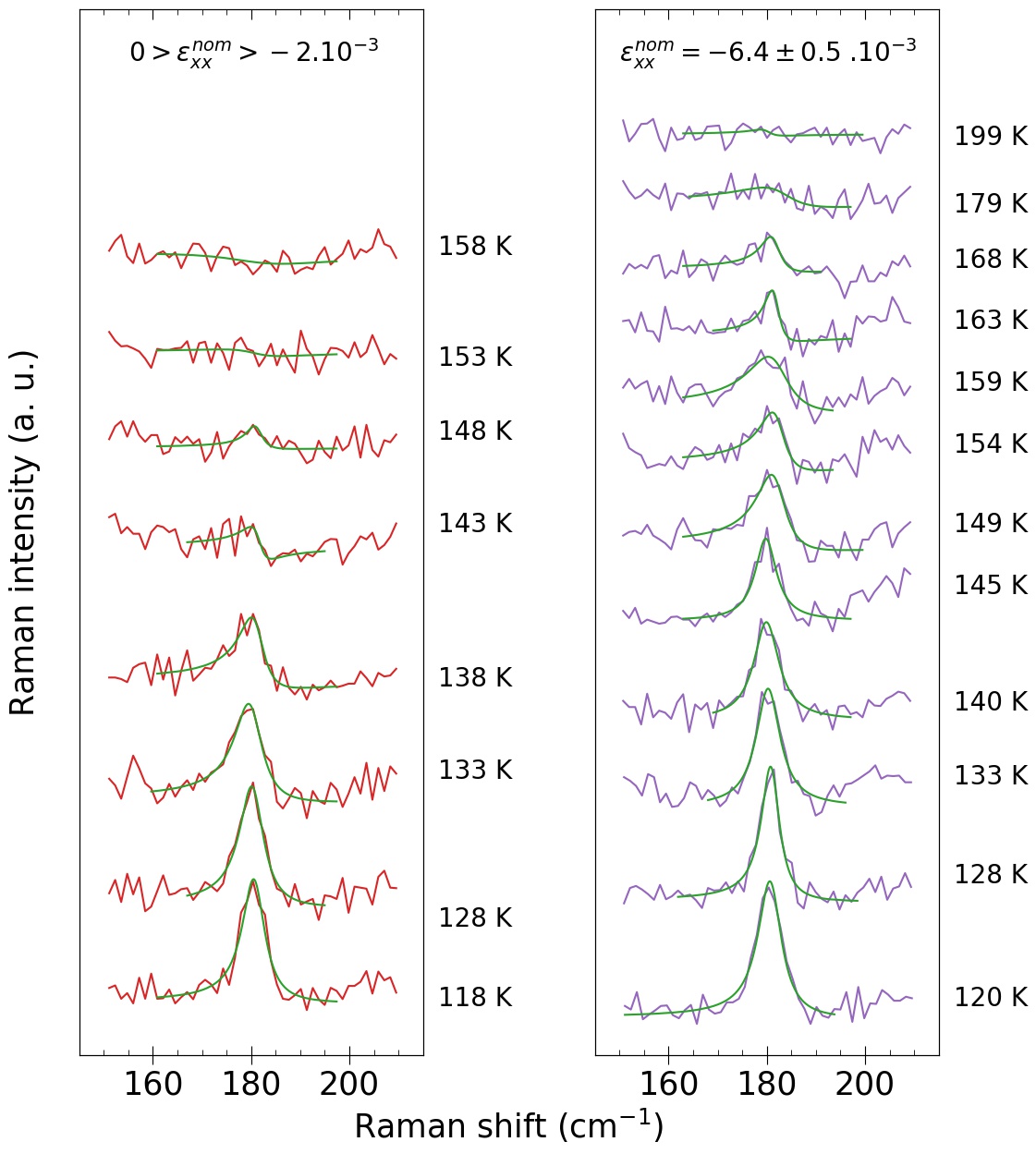}
\caption{Fano lineshape fitting of the As phonon as a function of temperature at two different applied strains}
\label{T-dep-fit}
\end{figure}

This expression was used to fit the As phonon as function of both strain and temperature in order to extract the integrated area of the bare phonon mode given by $\pi t_{ph}^2$. In order to fit the data it was necessary to add  an additional non-interacting background $y_0$. For spectra with a well-defined and intense peak such as at low temperature below $T_S$ and/or large strain, good convergence was found when keeping all parameters free. However for spectra displaying a weaker peak, we found it necessary to keep constant at least one of the parameters to ensure a more systematic fit convergence and we chose to keep $B=\frac{v}{t_e}$ constant. We checked that the fitted values of $\omega_0$ and $t_{ph}$ were only weakly dependent on different choices of $B$, typically within the standard error bars of the fits. The fits as a function of strain and temperature are shown in Fig. \ref{Strain-dep-fit} and \ref{T-dep-fit}. The extracted amplitude $t_{ph}$ was used for Fig. \ref{fig4} and \ref{fig5} of the main text.

\par
\begin{figure}
\includegraphics[width=0.45\textwidth]{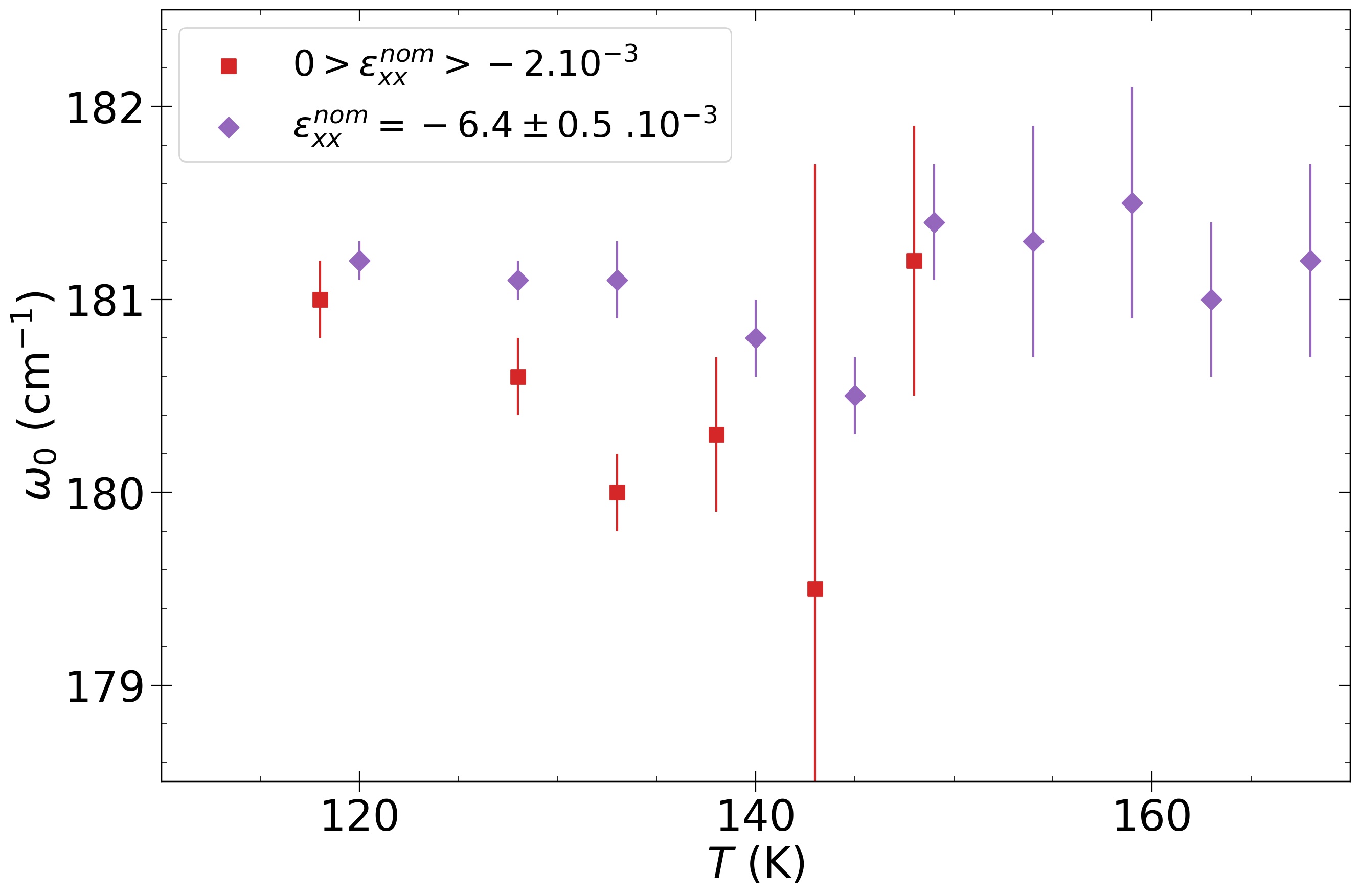}
\caption{\label{fano-energy} Temperature dependence of the effective phonon energy $\omega_0$ as extracted from the Fano lineshape fitting. At low strain the fitting above 148~K yielded very high error bars for $\omega_0$ due to the very weak phonon intensity and the corresponding values are not shown.}
\end{figure}

In the temperature dependence under strong strain (Fig.~\ref{T-dep-fit}), the fits also reveal a small anomaly of $\omega_0$ (about 1 cm$^{-1}$) at around 145~K (Fig.~\ref{fano-energy}). A similar softening was observed previously at zero strain at $T_{S,N}$  \cite{rahlenbeck_phonon_2009}.  We believe the observed softening in the high strain data is due to the SDW transition $T_N$ which remains well-defined even in the presence of strain and is only weakly affected by it, in agreement with NMR data of Kissikov et al. \cite{kissikov_uniaxial_2018}.

\section{Finite element simulation}
\label{Finite element simulation}

In this paper, we plotted the different quantities depending on the nominal strain in the $x$ direction $\epsilon_{xx}^{nom} = \frac{\delta L}{L_{0}}$, with $L_{0}$ the length of the suspended part of the sample when no stress is applied, and $\delta L$ the change of this length upon applied stress. $\epsilon_{xx}^{nom}$ is different from the nematic strain $\epsilon_{B_{2g}}$, with $\epsilon_{B_{2g}} = \frac{1}{2} (1 + \nu) \mu \epsilon_{xx}^{nom}$. $\nu$ is the Poisson ratio in the $ab$ plane and $\mu$ is the strain transmission coefficient through the epoxy glue. In BaFe$_2$As$_2$ the elastic coefficient corresponding to $B_{2g}$ strain, $C_{66}$, strongly softens upon cooling from 300~K to $T_{S/N} \approx 138$~K~\cite{yoshizawa_structural_2012}, which results in a strong decrease of the Young modulus in the $x$ direction~\cite{bohmer_nematic_2014}, and an increase of both $\mu$ and $\nu$. Therefore comparing the dependence of $\phi_{nem}^{As}$ on $\epsilon_{xx}^{nom}$ at different temperatures to evaluate the susceptibility can be misleading. To tackle this issue, we conducted finite element simulations to evaluate $\epsilon_{xx}$ and $\epsilon_{B_{2g}} = \epsilon_{x^{\prime}y^{\prime}}$. 
\par
\begin{figure}
    \centering
    \includegraphics[width=0.49\textwidth]{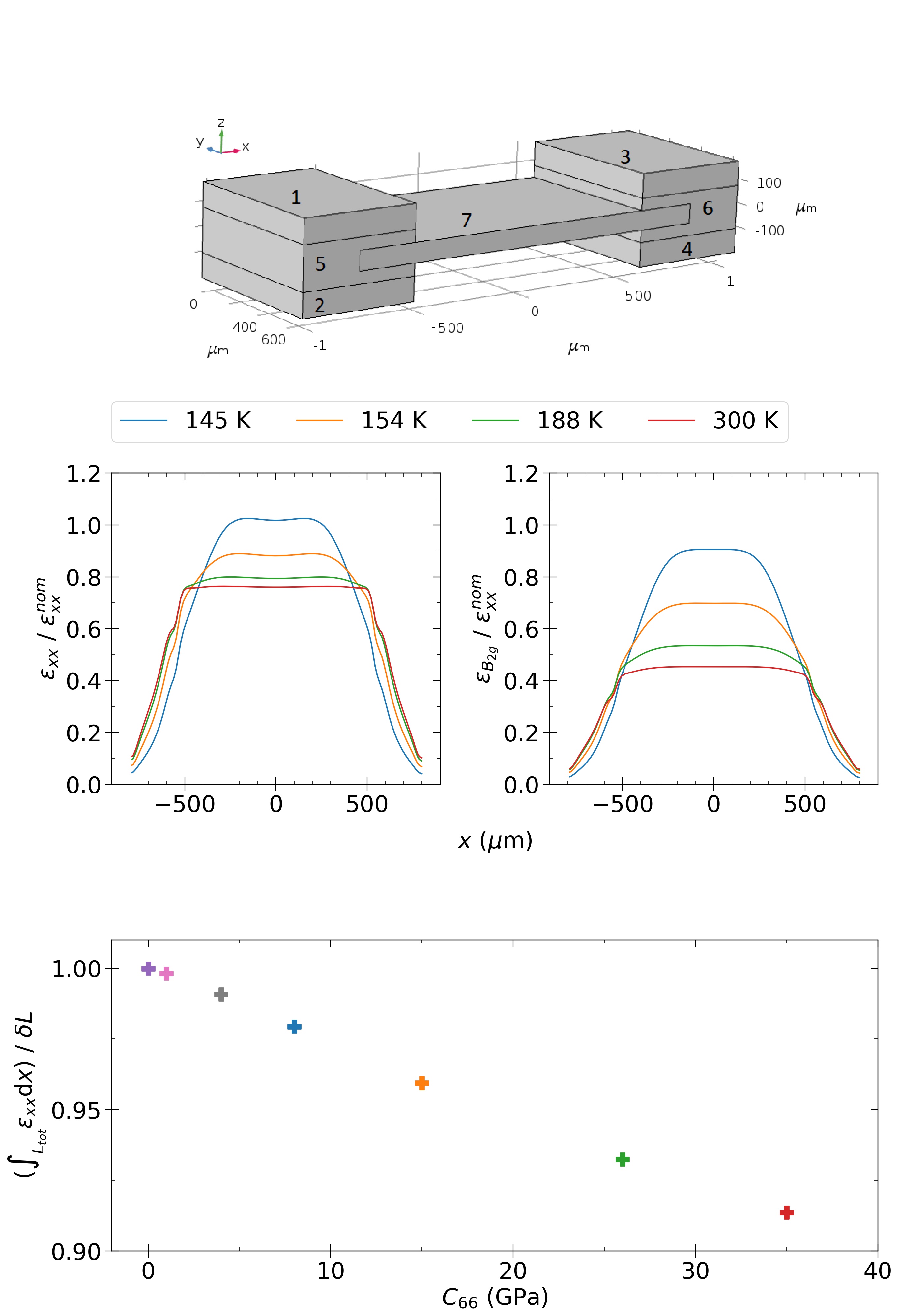}
    \caption{(a) Geometry of the built model. The domains correspond to : 1 to 4 : titanium plates ; 5 and 6 : epoxy glue ; 7 : sample. A displacement is given to domains 1 and 2, whereas domains 3 and 4 are held still. (b) Contribution of $\epsilon_{xx}$ and $\epsilon_{B_{2g}}$ in the total strain $\epsilon_{xx}^{nom}$. (c) Comparison between the maximum displacement in the sample and the applied displacement.}
    \label{figComsol}
\end{figure}

Fig.~\ref{figComsol}(a) shows the geometry of the model built for the simulation. We took 15~GPa and 0.3 respectively for the Young modulus and Poisson ratio of the Stycast 2850FT epoxy glue~\cite{barber_thesis_2018}, and 105~GPa and 0.33 for the titanium plates~\cite{ikeda_symmetric_2018}. On first approximation we considered these quantities to remain constant between 300~K and $T_{S/N}$, which probably leads to a slight overestimation of $\epsilon_{xx}$ and $\epsilon_{B_{2g}}$ at 300~K, as the epoxy glue is softer at high temperature. For BaFe$_2$As$_2$, we used the complete elastic tensor with the elasticity coefficients in the $x^{\prime}y^{\prime}z$ frame taken from~\cite{fujii_diverse_2018}. We extrapolated to 300~K the few elasticity coefficients not measured at room temperature by~\cite{fujii_diverse_2018}. In particular, we took for $C_{66}$ 8, 15, 26 and 35~GPa respectively for 145, 154, 188 and 300~K. We chose 20~$\upmu$m as the upper limit for the size of the finite elements in the sample domain, and some larger sizes in the epoxy glue and titanium domains.

The results for the simulation at the four temperatures for both $\epsilon_{xx}$ and $\epsilon_{B_{2g}}$ are plotted in Fig.~\ref{figComsol}(b). We observe that the strain is homogeneous in the middle region of the sample, where the laser spot is located, on a scale much larger than the spot size of $\sim 50$~ $\upmu$m. Fig.~\ref{figComsol}(b) shows results for the strain along an $x$ line in the bulk of the sample, but we checked that the strain displays a perfect homogeneity along the $z$ direction, and a sufficiently good one in the $y$ direction on the scale of the spot size. 

Because of the strong inhomogeneity of $\epsilon_{xx}$ along the length outside the middle region, with regions for $|x| > 500$~$\upmu$m being quite low strained, the resulting $\epsilon_{xx}$ in the middle region can exceed the applied strain $\epsilon_{xx}^{nom}$. This excess is weak at 145~K (3~\%), but strongly increases as we take $C_{66} \rightarrow 0$~GPa: the simulations give $\mu = $ 1.6/2.2/4.3 for $C_{66} = $ 1~GPa / 100~MPa / 0~GPa, with a simultaneous drastic reduction of the size of the middle region of homogeneity. This inhomogeneity enhancement might be problematic when probing the sample very close to $T_{S}$ but we note that in a real sample local defects will likely play an important role not captured by our simulation. 

In Fig.~\ref{figComsol}(c), we check that, for a given displacement $\delta L$, the maximum displacement in the sample, i.e. the displacement at the mobile end $u = \int_{L_{tot}}\epsilon_{xx} dx$, with $L_{tot}$ the total length of the sample (1600~$\upmu$m), is lower than $\delta L = \int_{L_{tot}}\epsilon_{xx}^{nom} dx$, even for $C_{66} \rightarrow 0$~GPa. This ratio can serve as a proxy to estimate the amount of strain in the epoxy glue.

In the following, we consider only the values of $\epsilon_{xx}$ and $\epsilon_{B_{2g}}$ in the middle region (at $x=0$~$\upmu$m in Fig.~\ref{figComsol}(b)). As expected, both quantities increase with decreasing $T$. In particular, $\frac{\epsilon_{B_{2g}}(145~\mathrm{K})}{\epsilon_{B_{2g}}(154~\mathrm{K})} \approx 1.3$. This significant ratio likely participates in the $\frac{\partial \phi_{nem}^{As}}{\partial \epsilon_{xx}^{nom}}$ difference between the two temperatures evidenced in Fig.~\ref{fig4}, but is not enough to explain it. Our simulations confirm that the change of slope is not merely an effect of the change of transmission, but $\chi^{As}_{nem}$ is quantitatively modified when approaching $T_{S/N}$ from above, thus supporting the view that it at least partly reflects the underlying electronic nematic susceptibility.

\section{Shear modulus under finite stress}
\label{Shear modulus under finite stress}
At vanishingly small stress the shear modulus softens to essentially zero at $T_S$ as observed in ultrasound and Young's modulus measurements \cite{yoshizawa_structural_2012,fernandes_effects_2010,bohmer_nematic_2014}. This softening can complicate the interpretation of temperature dependent data  under constant nominal strain $\epsilon_{xx}^{nom}$ reported in Fig. \ref{fig5} since its $B_{2g}$ component will depend significantly on temperature close to $T_S$ even if the nominal strain $\epsilon_{xx}^{nom}$ is kept constant \cite{sanchez_spontaneous_2020}. Published elastic measurements of $C_{66}$ can be used to evaluate both the Poisson ratio $\nu$ and the transmission coefficient and thus deduce the $B_{2g}$ component of the applied strain in the low strain limit. This is discussed in the main text and in Appendix~\ref{Finite element simulation}. Outside this regime however, we cannot rely on elastic measurements data. In particular, under strong stress the lattice is expected to stiffen due to the finite nematic order parameter, thus suppressing the observed softening and resulting in less temperature dependent Poisson ratio and transmission coefficient. Here we discuss this effect and its potential impact on the temperature dependent data under high-strain shown in Fig. \ref{fig5}. For this we need to include explicitly both electronic and lattice degrees of freedom in the Landau free energy. The Landau free energy of a coupled nematic-orthorhombic transition includes an electron nematic order parameter $\phi$ and a lattice distortion $\epsilon$ both belonging to the $B_{2g}$ representation of $D_{4h}$ point group. They are coupled bi-linearly via a nemato-elastic constant $\lambda > 0$. We assume that the system is under a stress $\sigma > 0$ whose $B_{2g}$ component is coupled linearly to $\epsilon$. Here we will limit ourselves to $\phi^4$ terms in the expansion:
\begin{equation}
F=\frac{r}{2} \phi^2 + \frac{u}{4}\phi^4+\frac{C_{66}^0}{2}\epsilon^2 - \lambda\phi\epsilon - \sigma\epsilon
\end{equation}
$C_{66}^0$ is the bare shear modulus which we will assume to be temperature independent and $r$ is the inverse of the nematic susceptibility which follows a mean-field like behavior $r=a(T-T_0)$. $T_0$ is the bare nematic phase transition in the absence of a coupling to the lattice. $\sigma$ and $\epsilon$ are the $B_{2g}$ component of the stress and strain tensor respectively.

\begin{figure}
\includegraphics[width=0.45\textwidth]{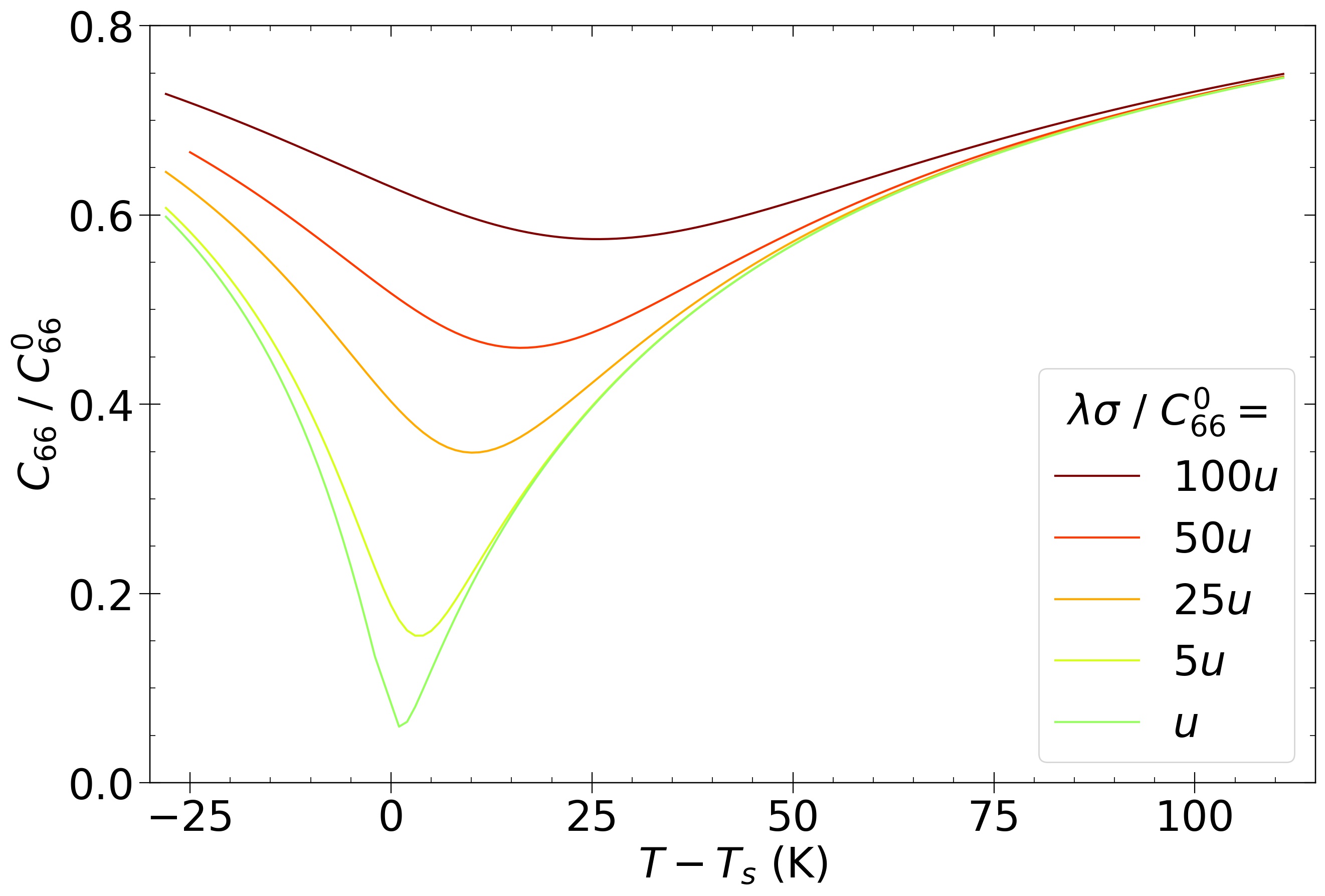}
\caption{\label{Cs-stress} Temperature dependence of the shear modulus $C_{66}$ under increasing $B_{2g}$ stress $\sigma$. Here we have used $T_S=138$~K, $T_0=100$~K, $u=1$, $a=1$ and $C^{0}_{66} = 30$~GPa}
\end{figure}

We first minimize the free energy with respect to $\phi$ and $\epsilon$ giving:
\begin{equation}
\sigma=C_{66}^0\epsilon-\lambda\phi
\end{equation}
\begin{equation}
\epsilon=\frac{r\phi+u\phi^3}{\lambda}
\end{equation}
The two equations can be combined to give an equation for the nematic order parameter under constant stress:
\begin{equation}
    (r-\frac{\lambda^2}{C_{66}^0})\phi +u\phi^3 -\frac{\lambda\sigma}{C_{66}^0}=0
    \label{OP-stress}
\end{equation}
Note that the nematic transition has been shifted to higher temperature $T_S$ due to the finite nemato-elastic coupling: $T_S=T_0+\frac{\lambda^2}{aC_{66}^0}$. 
The renormalized shear modulus $C_{66}$ due to nemato-elastic coupling can be computed using the partial derivatives of $\phi$:
\begin{equation}
C_{66}=\frac{\partial\phi}{\partial\epsilon}\frac{\partial\sigma}{\partial \phi}
\end{equation}
Using the minimization conditions the partial derivatives are given by:
\begin{equation}
\frac{\partial\epsilon}{\partial\phi}=\frac{1}{\lambda}(r+3u\phi^2)
\end{equation}
\begin{equation}
\frac{\partial\sigma}{\partial\phi}=C_{66}^0\frac{\partial\epsilon}{\partial\phi}-\lambda
\end{equation}
Inserting into the expression of $C_{66}$ we get:
\begin{equation}
C_{66}=C_{66}^0-\lambda\frac{\partial\phi}{\partial\epsilon}=C_{66}^0-\frac{\lambda^2}{r+3u\phi^2}
\label{shear-modulus-stress}
\end{equation}
Using equation \eqref{shear-modulus-stress} and the temperature dependence of the nematic order parameter under constant stress $\sigma$ \eqref{OP-stress}, we can plot the temperature dependence of the shear modulus under finite stress (Fig.~\ref{Cs-stress}). 

\par
For sufficiently strong stress the softening of the shear modulus $C_{66}$ is indeed strongly suppressed, and we expect the Poisson and transmission ratio to be much less temperature dependent. The question is then to evaluate to which regime the temperature dependent measurements under high-strain of Fig. \ref{fig5} correspond to. According to X-ray data under strain of Sanchez et al. \cite{sanchez_spontaneous_2020} on a Ba122 crystal with 3 percent Co doping, both the Poisson ratio and the transmission coefficient become essentially temperature independent for nominal applied strains $\epsilon_{xx}^{nom} > 5\times 10^{-3}$. Since the measurements of Fig. \ref{fig5} were performed at $\lvert \epsilon_{xx}^{nom} \rvert > 5.10^{-3}$, we are likely in the regime where the shear modulus is essentially temperature independent.
\par
From our temperature dependent Raman data under strong strain, we can also estimate in which regime we are. Indeed assuming the As phonon intensity in the $B_{2g}$ representation is a good proxy of the square of the nematic order parameter $\phi^2$, we see that the nematic order parameter under strong strain (above $\lvert\epsilon_{xx}^{nom}\rvert=4.10^{-3}$) at 145~K is very close (within 25~\%) to the saturated value at 118~K (Fig.~\ref{fig4} and \ref{fig5}). Using this information, we can infer the amount of shear modulus softening near $T_S$ (where it is strongest) using the expression of the shear modulus as a function of the nematic order parameter $\phi$.
If we assume that the nematic order parameter under stress at $T_S=138$~K is a fraction $\alpha^{1/2}$ of the low temperature (saturated) nematic order parameter at zero stress: 
\begin{equation}
\phi^2(T_S,\sigma\neq0)=\alpha\phi^2(T=0~\mathrm{K}, \sigma=0)=\frac{\alpha aT_S}{u}
\end{equation}
Substituting into the equation for the shear modulus \eqref{shear-modulus-stress}, we can estimate its value at $T_S$:
\begin{equation}
C_{66}(T_S)=C_{66}^0-\frac{\lambda^2}{a(T_S-T_0)+3aT_S\alpha}
\end{equation}
Using $\frac{\lambda^2}{a}=C_{66}^0(T_S-T_0)$, we obtain:
\begin{equation}
C_{66}(T_S)=C_{66}^0 \left( 1-\frac{T_S-T_0}{(1+3\alpha)T_S-T_0} \right)
\end{equation}
For Ba122 we have $T_S=138$~K and from shear modulus and Raman data (under zero strain) $T_0=100$~K \cite{gallais_charge_2016,yoshizawa_structural_2012}. Taking a conservative $\alpha=0.5$ (70~\% of the low temperature order parameter at $T=145~\mathrm{K}\sim T_S$ at high-strain) we obtain $C_{66}\sim 0.84 C_{66}^0$. This indicates that the complete softening of $C_{66}$ at zero strain is replaced by a much smaller softening of about 16~\% at high strain giving an upper bound on the temperature evolution of $C_{66}$. In turn a 16~\% change in $C_{66}$ will imply a change in the Poisson ratio $1+\nu$. To evaluate this we use the relationship between the $C_{66}$ and $1+\nu$: $1+\nu=\frac{2C}{C+C_{66}}$, with $C=\frac{1}{2}(C_{11}+C_{12})-\frac{C_{13}^2}{C_{33}}$. Except $C_{66}$ the elastic coefficients are weakly temperature dependent. Using $C_{11}=95$~GPa, $C_{12} = C_{13} = 17$~GPa, $C_{33}=75$~GPa, and $C_{66}=35$~GPa \cite{fujii_diverse_2018}, we obtain a change of 7~\% in $1+\nu$ between $T_S$ and high-temperatures. From the simulations (Appendix~\ref{Finite element simulation}), the change in the transmission ratio $\mu$ will be even less. To conclude, changes in the $B_{2g}$
component of the strain as a function of temperature are unlikely to significantly affect the observed temperature dependence of $\phi_{nem}^{As}$ under high-strain reported in Fig. \ref{fig5}.

\section{Nematic order parameter under a symmetry breaking field}
\label{Nematic order parameter under a symmetry breaking field}

The behavior of the nematic order parameter under a symmetry breaking field can be captured by a Landau free energy with an electron nematic order paramater $\phi$ and a symmetry breaking field $h$:
\begin{equation}
F=\frac{r}{2} \phi^2 + \frac{u}{4}\phi^4+\frac{w}{6}\phi^6 - h\phi
\end{equation}
In our case $h$ is a $B_{2g}$ field which couples linearly to $\phi$ and $r = a( T-T_S)$ ($T_S$ is the renormalized nematic transition temperature).
Minimizing the free energy with respect to $\phi$ we obtain:
\begin{equation}
r\phi+u\phi^3+w\phi^5-h=0
\label{Min1}
\end{equation}

\begin{figure}[]
\includegraphics[width=0.45\textwidth]{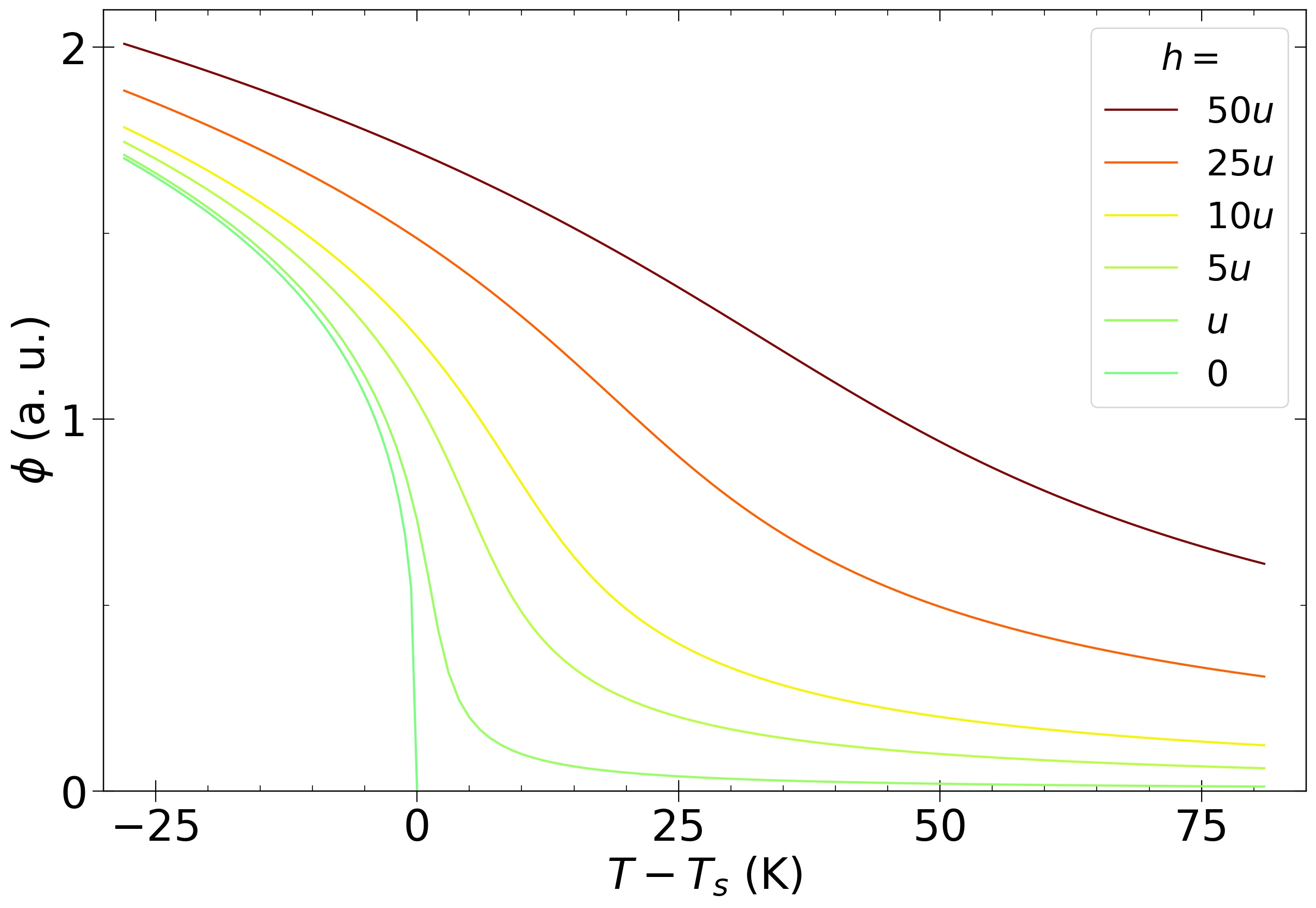}
\caption{\label{OP-field} Temperature dependence of the nematic order parameter $\phi$ under increasing symmetry breaking field $h$. Parameters used are ($a$,$u$,$w$)$=$(1,1,3).}
\end{figure}

The above equation was solved for $\phi$ numerically as a function of temperature $T$ and for various field $h$ keeping all other parameters constant (Fig.~\ref{fig5} and \ref{OP-field}). In Fig. \ref{fig5}, the parameters used were $h=5u$ and $h=25u$ for the low and high strain data respectively.

\end{document}